\documentclass[a4paper,reprint,showpacs,pra]{revtex4-1}
\usepackage{amsfonts,amsmath,epsfig}
\usepackage{bookmark}
\newcommand{\be}{\begin{equation}}
\newcommand{\ee}{\end{equation}}
\newcommand{\re}{\mbox{$\mathrm{Re}$\,}}
\newcommand{\im}{\mbox{$\mathrm{Im}$\,}}

\begin{document}
\title{Natural generalization of the ground-state Slater determinant to more
than one dimension}
\author{D.~K.~Sunko}
\email{dks@phy.hr}
\affiliation{Department of Physics, Faculty of Science, University of
Zagreb,\\ Bijeni\v cka cesta 32, HR-10000 Zagreb, Croatia.}

\begin{abstract}

The basic question is addressed, how the space dimension $d$ is encoded in the
Hilbert space of $N$ identical fermions. There appears a finite number
$N!^{d-1}$ of many-body wave functions, called shapes, which cannot be
generated by trivial combinatorial extension of the one-dimensional ones. A
general algorithm is given to list them all in terms of standard Slater
determinants. Conversely, excitations which can be induced from the
one-dimensional case are bosonised into a system of distinguishable bosons,
called Euler bosons, much like the electromagnetic field is quantized in terms
of photons distinguishable by their wave numbers. Their wave functions are
given explicitly in terms of elementary symmetric functions, reflecting the
fact that the fermion sign problem is trivial in one dimension. The shapes act
as vacua for the Euler bosons. They are the natural generalization of the
single-Slater-determinant form for the ground state to more than one
dimension. In terms of algebraic invariant theory, the shapes are
antisymmetric invariants which finitely generate the $N$-fermion Hilbert space
as a graded algebra over the ring of symmetric polynomials. Analogous results
hold for identical bosons.

\end{abstract}


\maketitle
\section{Introduction}

Quantum effects are sometimes counter-intuitive because physics
happens in the space of wave functions, not in the geometrical ``laboratory''
space of Newtonian mechanics. Conversely, molecular isomerism, the phenomenon
that a given set of identical atoms can arrange itself in molecules of
different shapes, is quite intuitive geometrically. The question arises, how
is it manifested in wave-function space. The discreteness of wave functions
must somehow limit the relative positions in laboratory space. In particular,
one would like to have a qualitative argument, which shapes are possible
solutions of the many-body Schr\"odinger equation, without a full calculation.

The choice of ground-state wave function --- i.e.\, particular shape among
possible isomers --- is evidently related to the choice of a wave function
with pronounced correlations, or collectivity. The notion of collectivity is
usually taken to mean that the energy cannot be expressed as the sum of
energies of single-particle wave functions. That intuition cannot be literally
true, because the Kohn-Sham theorem~\cite{Kohn65} shows that it is possible to
construct artificial single-particle states precisely by the requirement that
the exact ground state energy can be expressed in this way.

Since Dirac introduced them~\cite{Dirac26}, Slater
determinants~\cite{Slater29} have been the only fundamental antisymmetric
forms available to construct optimized wave functions. Being a complete basis
for the N-body Hilbert space, they encourage a functional-analytic,
essentially structureless, view of that space, as a vector space in which the
ground state is just one particular linear combination of Slater determinants
among many.

In particular, the Kohn-Sham method~\cite{Kohn65} is a special search in
coefficient space, constrained by the requirement that the final linear
combination can be written as a single Slater determinant in some new
single-particle wave functions. However, the restriction to a
Slater-determinant form is arbitrary, basically due to a lack of a priori
alternatives. The price paid for it is that the new single-particle wave
functions are artificial, even if the ground-state energy is correct. If the
restriction is relaxed, the lack of structure in the functional-analytic
approach (one set of $c$-numbers is in principle as good as another) leads to
the impression that there is an infinity of possibilities to choose from.

Motivated by the above considerations, the present work explores a related but
more qualitative idea of collectivity, based on wave-function properties
rather than energies. If the single-particle wave functions are separable in
the Cartesian coordinates of laboratory space, good candidates for many-body
collective states should not inherit this separability. Such states are
multidimensional in some non-trivial way, which is given a rigorous meaning
here.

The main result is that there exist precisely $N!^{d-1}$ antisymmetric forms,
called \emph{shapes}, which are the basic building blocks of any antisymmetric
$N$-body wave function in $d$ dimensions. This result is rooted in the
algebraic theory of invariants~\cite{Derksen02}, indeed it is expected in that
context (known as ``Hilbert's 14th problem''). However, it is unexpected to
physicists and chemists, who are trained in the functional-analytic rather
than algebraic approach to Hilbert spaces. In the former case, Hilbert space
is viewed as a vector space spanned by an infinity of Slater determinants
$\Psi_i$, in which any wave function may be written as
\be
\Psi=\sum_i c_i\Psi_i,
\ee
where the $c_i$ are $c$-numbers. When a vector space is endowed with an
additional vector multiplication operation, it becomes an algebra. When the
vectors are complex functions, the natural vector multiplication is just
ordinary multiplication of functions. The switch to the algebraic approach is
thus technically manifested as a generalization of the $c_i$ to symmetric
functions of the space coordinates. Then it turns out that the sum becomes
\emph{finite}: many-body Hilbert space is a finite-dimensional algebra. Only a
finite number of antisymmetric forms $\Psi_i$ is needed to generate the whole
Hilbert space, now viewed as a graded algebra over the ring of symmetric
polynomials (for $N$ finite). These forms $\Psi_i$ are just the generators of
the Hilbert-space algebra, called shapes here. In one dimension, there is only
one shape, which is the ground-state Slater determinant. Thus shapes are
generalizations of this particular Slater determinant to more than one
dimension. They can always be expressed as superpositions of Slater
determinants, because the latter are a complete basis. The connections between
the algebraic and functional-analytic points of view for the physical $N$-body
problem have only begun to be explored in the present work.

The shapes extend the notion of a vacuum state in an explicit and formally
rigorous sense: the usual ground-state energy shift $\exp\left(-\beta
E_{gs}\right)\equiv q^{E_{gs}}$ in the partition function is replaced by a
\emph{shape polynomial} $P(q)$, which counts all possible antisymmetric forms
which can play the role of a vacuum. It is the generating function of the
shapes. An algorithm is provided which generates the Hilbert-space span of all
shapes in terms of Slater determinants. It may be interpreted as a
machine-assisted way to generate \textit{ans\"atze} for correlated ground
states when $d>1$.

Excitations of any one of these vacuum states are described by the
symmetric-function coefficients, in other words they are bosonic. In contrast
to the shapes, they can be extended from $d=1$ as if the space directions were
color labels, combined in all possible ways. These symmetric states are called
\emph{Euler bosons}, because their partition function was first obtained by
Euler~\cite{Euler51}. Euler bosons cannot exist by themselves. Each wave
function in the scheme is based on some single shape, with or without an
arbitrary number of excitations (Euler bosons) on top of it. In brief, the
shapes represent all possible many-body vacua for the Euler bosons.

The classification in the present form does not include spin, and refers to
only one kind of particle. Neither is an essential limitation. Including spin
and different kinds of particles amounts to combining several generating
algorithms of the type introduced here multiplicatively, which is unnecessary
for an initial description. It is possible to think of the states here as
referring to a concrete system, such as entangled atoms~\cite{King06}, or
electrons in a quantum dot~\cite{Warburton98,Kalliakos09}, or in the vacancy
of an electride~\cite{Miao14}, but an important aspect of the results is their
abstraction and generality, based on a topological (node-counting)
classification of wave functions, which is universal. In fact the basic
algorithm operates at the level where single-particle wave functions are
represented by formal powers, so that a term like $t^k$ refers to Hermite
polynomials $H_k(x)$ in one realization and to standing waves $\sin(k+1)x$ in
another. No result depends on the particular realization.

The article consists of two parts. The first is a self-contained derivation of
all the results in an abstract setting. The basic counting result is
established, with a recursion for the shape polynomial. The one-dimensional
case is solved in terms of the Euler bosons. A polynomial deflation algorithm
is introduced to express the Euler-boson wave functions in terms of standard
Slater determinants. This algorithm is used in $d>1$ to represent all trivial
(separable) states in an ordered succession of subspaces, finding the shapes
as the remainder (orthogonal complement) at each level. The second part
consists of examples and illustrations. For $N=3$ particles in $d=2$
dimensions, all the six shapes are constructed step by step. Some numerical
experiments are performed with the Coulomb interaction, to check that the
scheme is not unstable with respect to it. Variational functions and
simulations are discussed, with a minimal example.

While all the main formulas refer to fermions, in general the results for
identical bosons are very similar. This may have direct repercussions for
systems of entangled atoms. The bosonic case is compared to the fermionic one
at the end, before the discussion and conclusions. Some textbook mathematics
is collected in the appendices to make the article better self-contained.

\section{The basic counting result}

\subsection{Partition function}

In physics, the partition function, or sum over states, is typically used in
the context of thermodynamics, with the idea that each ``state'' being counted
is thermodynamically possible, in the sense that it is an energy level of the
actual system under consideration. In the present work, a more general
approach is taken, where a ``state'' is simply any wave function, irrespective
of whether there exists a Hamiltonian of which it is an eigenfunction. The
only requirement on the partition function is that it count the states
faithfully, i.e.\ each distinct wave function should appear exactly once.

If the single-particle wave functions are separable in Cartesian coordinates,
there is a natural organizing principle for counting all states. Each
many-body Slater determinant built out of such single-particle wave functions
has some number of single-particle nodes in each direction in space, say
$n_x$, $n_y$, $n_z$, for $d=3$. The list of all Slater determinants with a
given total number of nodes $E\equiv n_x+n_y+n_z$ is evidently finite.
Increasing the total number of nodes one by one, all possible wave functions
appear exactly once, so they can be counted faithfully.

The above scheme introduces the important notion of \emph{grading}, which is
just counting the total number of nodes $E$. All $N$-body wave functions
spanned by Slater determinants of the same grade $E$ form a closed subspace of
the Hilbert space, because a linear combination of such functions is itself a
wave function of the same grade.

Clearly, one realization of this scheme is the familiar harmonic-oscillator
well, for which the grade $E$ is also the energy, so that the graded states
are simultaneously energy eigenstates, and the sum over states, organized by
grade, also has the usual thermodynamic meaning. Although the
harmonic-oscillator picture is very useful for the visualization of various
results, it should not be construed that they are valid only for the
oscillator. Even the limitation to separable single-particle wave functions is
not strictly necessary. It is retained throughout this article to fix ideas,
because it easily produces explicit formulas. The main result is an intrinsic
property of $N$-body Hilbert space, because the dimension of an algebra (the
number of its generators) does not depend on any particular realization.

In order to implement the main idea from the Introduction, one should see how
many $N$-body wave functions in $d$ dimensions one can obtain which are
separable across the space dimensions. If the $N$-body functions in one
dimension are counted by some partition function, call it $Z_E$, then all
separable states are counted by $\left(Z_E\right)^d$. These are explicitly
constructed by labeling the space directions with different colors, and
combining the corresponding 1D wave functions in all possible ways. It follows
from this interpretation that these states are not all that can be found when
$d>1$. The reason is that the Pauli principle operates only upon the exchange
of the full (vector) coordinates of a pair of particles, while the iterated
$d=1$ states impose the antisymmetrization for each axis (coordinate
projection) individually, so that they are too restrictive when $d>1$.

Therefore the essential idea of the present classification is to write the
partition function of the $d$-dimensional system of $N$ identical fermions as
\be
Z_d=\left(Z_E\right)^dP_d(N,q).
\label{Zdfact}
\ee
For $d=1$ the \emph{ansatz} reduces to
\be
Z_1=Z_E P_1(N,q)\equiv Z_E q^{{E_{gs}}(N)},\quad q=e^{-\beta\epsilon},
\label{ZE}
\ee
where $E_{gs}(N)$ is the ground-state energy in units of $\epsilon$. The
``extra'' states allowed by the Pauli principle for $d>1$ are counted by the
factor $P_d$, which reduces in one dimension to a single monomial, the
``energy shift'' which counts the nodes of the ground state wave function.
These ``extra'' states are called shapes, and the term $P_d$ which counts them
turns out to be a polynomial for $d>1$, called the \emph{shape polynomial}.
For a graded counting scheme (harmonic well), $Z_E$ was first obtained by
Euler~\cite{Euler51}, hence the index $E$. ($Z_E$ is the same for bosons and
fermions, only $E_{gs}$ is different~\cite{Euler51,Andrews76}.) It will be
shown now that the total number of shapes is finite, $P_d(N,q=1)=N!^{d-1}$,
independently of any particular counting scheme, which proves that $P_d$ as
defined above is always a polynomial.

\subsection{High-temperature limit}\label{hightemp}

The non-interacting partition function for $N$ fermions in $d$ dimensions
obeys the well-known recursion relation~\cite{Bruno81,Ford71}
\begin{equation}
Z_d(N,\beta)=\frac{1}{N}\sum_{m=1}^N(-1)^{m+1}z_d(m\beta)Z_d(N-m,\beta).
\label{recz}
\end{equation}
Here $z_d(\beta)\equiv Z_d(1,\beta)$ is the one-particle partition function,
while $Z_d(0,\beta)\equiv 1$. In the infinite-temperature limit $\beta\to 0$,
or $q\to 1$, the term $m=1$ dominates the sum on the right, because the factor
$z_d(m\beta)$ is then the same for all values of $m$, while the factor
$Z_d(N-m,\beta)$ for $m=1$ strongly dominates those with $N-2$ and less
particles, when the temperature is high. Inserting the \emph{ansatz}
\eqref{Zdfact}, one gets ($Z_E=Z_1$ at $q=1$)
\begin{multline}
NZ_1(N,0)^dP_d(N,1)\\
=z_d(0)Z_1(N-1,0)^dP_d(N-1,1).
\end{multline}
At $\beta=0$, $Z_1(N,0)=z_1(0)^N/N!$ (classical limit with Boltzmann counting),
so that
\begin{multline}
Nz_1(0)^{Nd}P_d(N,1)/N!^d\\
=z_d(0)z_1(0)^{d(N-1)}P_d(N-1,1)/(N-1)!^d.
\end{multline}
Because the kinetic energy is additive in the space dimensions, we have
$z_d(0)=z_1(0)^d$, so that finally
\be
P_d(N,1)=P_d(N-1,1)N^{d-1},
\ee
which gives
\be
P_d(N,1)=N!^{d-1},
\label{asymptot}
\ee
as advertised in the Introduction. This result is general and exact, because
any system is a gas at sufficiently high temperature. Taking logarithms, it
means that the non-trivial states (shapes) have an extensive but finite
contribution to the free energy, which \emph{saturates} at sufficiently high
temperature. Because the number of shapes is finite, $P_d$ is a polynomial.

\subsection{The shape polynomial}

The above asymptotic result has been obtained without reference to any
particular counting scheme, or even one-body separability: there are always
$N!^{d-1}$ many-body wave functions which cannot be induced from the
one-dimensional ones. Among all general counting schemes, the grading scheme
is distinguished by the partition function being explicitly solvable. In one
dimension, the sum over fermion states as counted by nodes is~\cite{Andrews76}
\begin{multline}
Z_1=\sum_{0\le n_1<\ldots<n_N<\infty}q^{n_1+\ldots +n_N}
\\=
q^{N(N-1)/2}\frac{1}{1-q}\cdot\frac{1}{1-q^2}\cdots\frac{1}{1-q^N},
\label{Z1}
\end{multline}
a result due to Euler~\cite{Euler51}. (In the harmonic oscillator
interpretation, this result counts the states of $N$ fermions in a 1D
oscillator well, with $\varepsilon_n=n\hbar\omega$ and
$q=e^{-\beta\hbar\omega}$.)

Comparing Eqs. \eqref{ZE} and \eqref{Z1} identifies the 1D
partition function $Z_E$~\cite{Euler51}. Hence the \emph{ansatz}
\eqref{Zdfact} for the $d$-dimensional partition function reads
\be
Z_d(N,\beta)=P_d(N,q)\left(\prod_{k=1}^N\frac{1}{1-q^k}\right)^d.
\label{harmansatz}
\ee
The $d$-th power gives the trivial extension of the 1D case to $d$ dimensions.
The extra term $P_d(N,q)$ can now be calculated explicitly, proving that the
\emph{ansatz} is solvable when $Z_E$ counts wave functions by grade.

Inserting the \emph{ansatz} \eqref{harmansatz} into the recursion \eqref{recz}
gives (this formula was first derived by D.~Svrtan)
\be
NP_d(N,q)=\sum_{k=1}^{N}(-1)^{k+1}\left[C^{N}_{k}(q)\right]^dP_d(N-k,q),
\label{recurPd}
\ee
where $P_d(0,q)=P_d(1,q)=1$, and
\be
C^N_k(q)=\frac{(1-q^N)\cdots (1-q^{N-k+1})}{(1-q^k)}
\ee
is a polynomial, because $k$ always divides one of the $k$ successive powers
of $q$ in the numerator. Therefore $P_d(N,q)$ is also a polynomial, as
expected for a generating function of a finite number of states. For the
oscillator, the degree of this polynomial is the energy cutoff above which the
shapes saturate.

Significantly, the recursion \eqref{recurPd} includes the ground-state shift
$E_0$ into the degree of the polynomial $P_d(N,q)$, which provides formal
reason to claim that the \emph{ground state is a shape}. This carries over
even to $d=1$, where $P_d$ consists of a single term. The recursion finds the
ground-state number of nodes (``energy'') as the lowest power of the
polynomial $P_d(q)$.

\section{Bosonisation of the 1D Fermi gas}

\subsection{Euler bosons}

Apart from the finite shift (``ground-state energy'') $N(N-1)/2$, the
remaining terms in Eq. \eqref{Z1} may be interpreted as the appearance of $N$
harmonic oscillators, mutually distinguishable, each having a different energy
spacing, $\hbar\omega_k=k\hbar\omega$, $k=1,\ldots,N$, but without a
zero-point energy of their own. The principal purpose of this section is to
obtain the wave functions of these \emph{Euler bosons}.

The factored form of Eq. \eqref{Z1} suggests that an arbitrary excited state
consists of two independent parts, so that its wave function may be factored as
well,
\be
\Psi\equiv\Phi\Psi_0,
\label{psifact}
\ee
where $\Psi_0$ is the ground-state Slater determinant, and $\Psi$ an arbitrary
one, describing an excitation in terms of some single-particle functions
$\phi_k(x)$. Here $\Psi_0$ accounts for the ground-state shift, while $\Phi$
is a \emph{symmetric} function in the $N$ variables, defined above as the
ratio of the two Slater determinants. The principal observation now is that
any 1D Slater determinant $\Psi$ is divisible by the ground-state determinant
$\Psi_0$, therefore $\Phi$ is a concrete symmetric polynomial, so that
Eq.~\eqref{psifact} is not just a notational trick. Namely, the 1D
single-particle wave functions consist of three parts,
\be
\phi_k(x)=N_kp_k(x)g(x),
\ee
a norm $N_k$ which depends only on the quantum number, but not on the
variable, an orthogonal polynomial $p_k(x)$ which depends on both, and
possibly a localization (e.g.\ exponential) term $g(x)$ which depends on the
variable but not on the quantum numbers. The norm and localization terms can
be factored out from the Slater determinants, because these have the same
quantum numbers in each row, and the same variable in each column. These terms
cross out in the numerator and denominator, up to a trivial overall factor.
Therefore the only parts remaining in the determinants themselves are the
orthogonal polynomials.

A Slater determinant of polynomials is itself a polynomial. It vanishes
whenever any two variables are equal, $x_i=x_j$ for $i\neq j$. By the
fundamental theorem of algebra, it must contain a term $(x_i-x_j)$ in its root
factorization for all pairs $i\neq j$. The denominator $\Psi_0$ contains all
these terms to lowest order, because the ground state has the smallest number
of nodes. Hence it divides the numerator $\Psi$. [The same conclusion applies
when $p_k(x)$ are trigonometric functions, which are algebraically just
shifted polynomials, $\cos kx\leftrightarrow u^k+u^{-k}$.]

The above reasoning is reduced to its essence if each single-particle wave
function is replaced by a symbolic power counting the number of nodes,
\be
\phi_k(x_i)\rightarrow t_i^k,\quad i=1,\ldots,N.
\label{map}
\ee
In this form it appears in mathematics textbooks, which leave the ``general''
polynomial case as an exercise, for the reader to be convinced that it brings
nothing new~\cite{Aitken39}. The denominator $\Psi_0$ then becomes the
well-known Vandermonde determinant~\cite{Aitken39},
\be
\Psi_0=\Delta(t_1,\ldots,t_N)\equiv
\prod_{1\le i<j\le N} (t_i-t_j).
\label{vandermonde}
\ee

This symbolic-power representation is the level of abstraction which we adopt
now. One can always specialize to the single-particle wave functions for a
particular problem by a reverse of the same mapping, the important point being
that it preserves the grading. The scheme works because it encodes the
essential behavior of nodes under multiplication and addition of functions. If
two functions are multiplied, the number of nodes is added. If the functions
are added, the number of nodes stays the same as that of the function with the
larger number of nodes. Pure powers behave in exactly the same way. In this
abstract representation, the ratio $\Phi$ is called the Schur
function~\cite{Stanley99} (see the Appendix). The physical statement that the
Slater determinants span the whole Hilbert space is mirrored by the statement
that the Schur functions are a complete basis for the symmetric polynomials.

From a physical point of view, the new insight is that there is ``really''
only one antisymmetric many-body function in one dimension, and that is the
ground-state Slater determinant:
\be
\sum_ic_i\Psi_i=\left(\sum_ic_i\Phi_i\right)\Psi_0,
\label{psi1d}
\ee
where $\Psi_i$ are arbitrary Slater determinants, and $\Phi_i$ are the
corresponding Schur functions. Clearly the term in parentheses is a
\emph{bosonic} wave function. The factored form reflects the factored sum over
states \eqref{Z1}, where $\Psi_0$ accounts for the constant term
$q^{N(N-1)/2}$, corresponding to the ground-state. The remaining question is,
which wave functions correspond to the geometric series in the other factor of
the partition function. These are the wave functions of the Euler bosons,
which describe all possible excitations.

One can guess the correct abstract form by considering the harmonic oscillator,
because for the latter there exists an explicit realization of the
mapping \eqref{map}. It is the Bargmann
transform~\cite{Bargmann61,Bargmann67}, which carries Hermite functions
$\psi_n(x)$ of a real variable $x$ into powers of a complex variable $t$:
\be
\mathcal{B}[\psi_n](t)=\frac{1}{\pi^{1/4}}
\int_{\mathbb{R}}
e^{-\frac{t^2+x^2}{2}+xt\sqrt{2}}\psi_n(x)
=\frac{t^n}{\sqrt{n!}}.
\ee
The Bargmann-transformed oscillator Hamiltonian is then
\be
H=\sum_{i=1}^N(t_i\partial_{t_i}+1/2)\hbar\omega.
\ee
Inserting the decomposition \eqref{psifact} into the Schr\"odinger equation
$H\Psi=E\Psi$, the equation for $\Phi$ becomes
\be
\sum_{i=1}^N(t_i\partial_{t_i})\Phi=\frac{E-E_0}{\hbar\omega}\Phi,
\label{Phi1}
\ee
which is clearly solved by any homogeneous polynomial in the $t_i$. Notice how
the zero-point term from $\hbar\omega/2$ has been absorbed into $E_0$, i.e.\
the left-hand side lacks the usual $1/2$. This equation is
``first-quantized,'' because the requirement that $\Phi$ be symmetric in the
$t_i$ must be added extraneously. To obtain a primitive realization of the
symmetry requirement (``second quantization''), invoke the change of variables
\begin{align}
e_1&=t_1+\ldots+t_N=\sum_it_i,\nonumber\\
e_2&=t_1t_2+t_1t_3+\ldots+t_{N-1}t_N=\sum_{i<j}t_it_j,\label{ek}\\
&\vdots\nonumber\\
e_N&=t_1t_2\cdots t_N.\nonumber
\end{align}
The $e_k$ are just the elementary symmetric functions, appearing e.g.\ in the
Vi\`ete formulas for the coefficients of the polynomial
$(t-t_1)\cdots(t-t_N)$, whose roots are the $t_i$. As is well known, the
transformation from roots to coefficients is regular as long as all $t_i\neq
t_j$. Its Jacobian in Bargmann space (volume element $d\re t_id\im t_i$) is
the square of the Vandermonde determinant \eqref{vandermonde}, $J=|\Delta|^2$.

All symmetric functions in the $t_i$ can be rewritten in the $e_k$. The $e_k$
are eigenfunctions of Eq. \eqref{Phi1} with eigenvalue $k$, so the Hamiltonian
is transformed to the $e_k$ basis as
\be
\sum_{i=1}^N(t_i\partial_{t_i})=\sum_{k=1}^Nk(e_k\partial_{e_k}),
\ee
whose eigenfunctions are all the monomials $e_1^{n_1}\cdots e_N^{n_N}$, with
eigenvalue $n_1+2n_2+\ldots +Nn_N$, and no symmetry restrictions: the $e_i$
are therefore \emph{distinguishable}, as implied by Euler's factorization in
Eq. \eqref{Z1}. Hence functions of the $e_k$ are a second-quantized
representation for the original many-body fermionic excitations, yet the
representation is purely bosonic. The $e_k$ are a complete basis for the
symmetric functions, and all their monomials are generated by the formal
expression
\be
\frac{1}{1-e_1}\cdot\frac{1}{1-e_2}\cdots\frac{1}{1-e_N}.
\label{eulergen}
\ee
Because $e_k$ has the eigenvalue $k$, substituting $e_k=q^k$ in the above
expression will give the corresponding canonical partition function,
recovering Euler's result. Obviously, the monomial $e_1^{n_1}e_2^{n_2}\ldots
e_N^{n_N}$ is the wave function of $n_1$ Euler bosons of type $1$, $n_2$ of
type $2$, etc. This identification is the main result of the present section.
It obviously carries over to the formal-power representation, again because
the energy in the oscillator case is the grading, or polynomial degree, in the
general case.

Two things have been accomplished by identifying the Euler bosons. The most
important one is finding a generating function for their wave functions, Eq.
\eqref{eulergen}, which corresponds precisely to the sum over states which
counts them, Eq. \eqref{Z1}. This will enable ``lifting'' the present result
to $d$ dimensions by way of Eq. \eqref{harmansatz} and thus identifying the
wave functions of the shapes, counted by $P_d(N,q)$, which is the main purpose
of the present article.

The other is a more qualitative development: Euler boson excitations have
direct physical connotations. Namely, the transformation \eqref{ek} is
non-linear, progressing from a pure sum to a pure product. In physics, product
wave functions correspond to (non-interacting) gases, while sum wave functions
are typically used as trial wave functions for liquids. In other words, the
progression from $e_1$ to $e_N$ is physically in terms of decreasing
collectivity: the lowest-grade Euler boson $e_1$ is the most collective
(liquid-like), while the highest-grade $e_N$ is least collective (most
gas-like). This simplicity of physical interpretation pleasantly reflects
their mathematical simplicity, because of which they may be readily calculated
by Vi\`ete's interpretation above, deserving the name \emph{elementary}
symmetric functions. By contrast, Schur functions are sophisticated
combinatorial objects. The most efficient prescription for their calculation
is to interpret them as generating functions of semi-standard Young tableaux,
which is quite a surprising insight~\cite{Stanley99} (see the Appendix). There
is no simple physical interpretation of this property, accounting perhaps for
the fact that representations of collective states in terms of Slater
determinants are rarely physically transparent.

\subsection{Deflation algorithm}

In the previous section, it was found that Euler bosons are the natural basis
of graded one-dimensional $N$-fermion wave functions. Slater determinants are
in a sense redundant: only one Slater determinant, the ground state, is
sufficient to generate the whole Hilbert space, with excitations described in
terms of Euler bosons. In order to generalize this result to more than one
dimension, it is necessary to obtain the Euler-boson wave functions
explicitly, in terms of standard Slater determinants.

First one must deal with a slight complication. Compare the wave
functions $e_2\Psi_0$ and $e_1^2\Psi_0$, say. Because
$e_1^2=(t_1+t_2+\ldots)^2$ contains terms like $t_1t_2$, which also appear in
$e_2=t_1t_2+\ldots$, the two wave functions will not be orthogonal. It is much
better to interpret the powers $e_i^k$ appearing in Eq. \eqref{eulergen} by
raising individual monomials in them to the required power without cross
terms, e.g.\ $e_1^2\to t_1^2+t_2^2+\ldots$, or in general:
\be
e_m^k\to \sum_{1\le i_1<\ldots<i_m\le N}\left(t_{i_1}\cdots t_{i_m}\right)^k,
\label{pwr}
\ee
which clearly keeps the terms orthogonal, because now no monomial appears
twice in the various geometric series. [Technically Eq. \eqref{pwr} is a
composition, or plethysm, of the $e_m$ and power sums $p_k$~\cite{Stanley99}.]

The deflation algorithm operates as follows. Take any monomial wave function
containing Euler bosons, e.g.\ $e_1e_2^2\Psi_0$. By power counting, this state
belongs to the fifth-excited ``oscillator'' level above the ground state.
Expand it as a polynomial in the formal variables $t_i$. All Slater
determinants in the fifth level can be similarly expressed as polynomials in
the same $t_i$. Now it is simply a matter of ordering the polynomial terms in
some definite (say lexicographic) order, to see which Slater determinant
contains the leading order monomial of the given polynomial wave function, and
subtracting it with the appropriate coefficient. Then the leading power of the
remainder is determined, and subtracted in the same way. Because the Slater
determinants are a complete orthogonal basis for each level, this procedure is
guaranteed to terminate.

In fact the procedure is redundant. The Slater determinants can themselves be
factored as in Eq. \eqref{psifact}, so there is no need to multiply out the
term $\Psi_0$. The problem boils down to expressing a given product of sums
like \eqref{pwr} in Schur functions, which is just a basis transformation
among symmetric functions. The reason for stating the algorithm in the
less efficient formulation is that it then generalizes directly to several
dimensions, where the analogous generalization of Schur functions is not
available.

It is essential for the deflation algorithm that one deal with Slater
determinants of \emph{unnormalized} single-particle states. In practice, this
means using formal powers $t_i^k$ , instead of $t_i^k/\sqrt{k!}$ as in
Bargmann space. All superpositions of Slater determinants are obtained among
such unnormalized determinants, and normalized as superpositions only after
being mapped back to some concrete realization. This will become clear in the
example in the second part of the paper.

The above technical considerations reflect a change of viewpoint. The
deflation algorithm in the algebraic approach corresponds to taking
projections in the standard functional-analytic approach. The algebraic
approach, chosen by the mapping \eqref{map}, brings one to consider
$N$-fermion Hilbert space as a space of antisymmetric polynomials, graded by
their degree. In one dimension, this space maps straightforwardly on the space
of symmetric polynomials, which is one way to understand why the fermion sign
problem~\cite{Loh90} is trivial when $d=1$. While this insight is undoubtedly
interesting, the true advantage of the algebraic approach appears in more than
one dimension. There it uncovers a fundamental structure of many-body Hilbert
space which is hidden in the functional-analytic approach, as will become
apparent in Sect.~\ref{genalg} below.

\section{The multidimensional case}

\subsection{Slater determinants}

A Slater-determinant state is obtained by denoting single-particle wave
functions as formal powers in $d$-plets of variables for each particle, say
the triplet $t,u,v$ for $d=3$. Then a general (unnormalized) Slater
determinant is written e.g.\ for $N=2$ particles
\begin{multline}
|\vec{m}_1,\vec{m}_2|=
|t_{1}^{m_{11}}u_{1}^{m_{12}}v_{1}^{m_{13}},
t_{2}^{m_{21}}u_{2}^{m_{22}}v_{2}^{m_{23}}|
\\=
\left|\begin{matrix}
t_{1}^{m_{11}}u_{1}^{m_{12}}v_{1}^{m_{13}}&
t_{2}^{m_{11}}u_{2}^{m_{12}}v_{2}^{m_{13}}\\
t_{1}^{m_{21}}u_{1}^{m_{22}}v_{1}^{m_{23}}&
t_{2}^{m_{21}}u_{2}^{m_{22}}v_{2}^{m_{23}}
\end{matrix}
\right|.
\label{slaterdef}
\end{multline}
An absolute ordering (e.g.\ lexicographic) on the $d$-dimensional integer
vectors $\vec{m}_i$, $i=1,\ldots,N$, is required to fix the phase of the
Slater determinants, which need not be explicit here. An example, to be used
later, is the ground state of $N=3$ particles in $d=2$ dimensions:
\be
g_0\equiv |(1,0),(0,1),(0,0)|=|t_1,u_2,1|=
\left|\begin{matrix}
t_{1}&t_{2}&t_{3}\\
u_{1}&u_{2}&u_{3}\\
1&1&1
\end{matrix}
\right|.
\label{g0}
\ee

\subsection{General algorithm for shape wave functions}\label{genalg}

As noted before, the factorized form of the wave function \eqref{psi1d}
follows the factorization \eqref{Z1} of the sum over states. Similarly the
$d$-dimensional sum over states \eqref{harmansatz} implies
the general form~\cite{Derksen02}
\be
\sum_{i=1}^{P_d(N,q=1)} \Phi_i\Psi_i
\label{genspan}
\ee
for wave functions, where the $\Phi_i$ are $d$-dimensional Euler boson states,
and the $\Psi_i$ are all the $P_d(N,q=1)$ states counted by the factor
$P_d(N,q)$ in the expression \eqref{harmansatz}. These states have been called
shapes above. Clearly they generalize the ground-state Slater determinant in
Eq. \eqref{psi1d} in such a way that an arbitrary wave function can be
expressed in terms of shapes, which must be antisymmetric wave functions,
because the $\Phi_i$ are symmetric. In Sect.~\ref{hightemp}, it has been
proven completely generally that $P_d(N,q=1)=N!^{d-1}$, therefore the whole
$N$-fermion Hilbert space can be \emph{finitely} generated with the shapes as
basic antisymmetric building blocks. They are generators of the
finite-dimensional Hilbert-space algebra induced by wave-function
multiplication. This is the main result of the present work, announced in the
Introduction. The purpose of this section is to generate the shape basis
explicitly, amounting to a constructive proof of the same result.

One can combine the deflation algorithm with the $d$-dimensional extension of
Eq. \eqref{eulergen} to obtain all the shapes. All possible Euler boson wave
functions in $d$ dimensions are obtained simply by multiplying $d$ copies of
Eq. \eqref{eulergen}, one for each set of variables $t_i,u_i,\ldots$,
representing the directions in space:
\be
\left[\frac{1}{1-e_1(t)}\cdots\frac{1}{1-e_N(t)}
\right]
\left[\frac{1}{1-e_1(u)}\cdots\frac{1}{1-e_N(u)}
\right]\cdots,
\label{generating}
\ee
where e.g.\ $e_1(u)=u_1+\ldots +u_N$, and so on. There is no similar closed
generating formula for the wave functions of the shapes, which would
analogously correspond to the shape-polynomial factor $P_d(N,q)$ in the
counting expression \eqref{harmansatz}. Instead we resort to the following
constructive algorithm.

Start with shapes at zeroth level, which are just the Slater determinants
spanning the (possibly degenerate) ground state level, which contains no Euler
bosons. Excite them by multiplying them with Euler bosons, noting that there
are only $d$ Euler bosons which carry one quantum of excitation, namely the
$e_1$ monomials, one for each direction in space. Multiplying the ground
state(s) with them gives all the states containing Euler bosons (``trivial
states'' for short) at the first excited level, so if it contains more than
$d$ states, the remainder (orthogonal complement) are the shapes at first
level. After applying the deflation algorithm to find the span of the trivial
states in terms of Slater determinants, a standard algebraic algorithm is
invoked to find the orthogonal complement of this vector space. The dimension
of the complement space is given by the corresponding coefficient in the shape
polynomial $P_d(N,q)$, which is a useful check on the implementation. Now one
iterates the procedure, multiplying the first-level shapes with the Euler
bosons $e_1$, and adding the ground states multiplied by all two-quanta
bosons, like $e_2$ and $e_1^2$, to obtain all the trivial states in the second
level. The span of the so-generated second-level trivial space is again found
by the deflation algorithm. The second-level shapes are the orthogonal
complement to that space, and so on until all shapes predicted by the shape
polynomial are found. In this way, the algorithm finds the Hilbert space span
of the shapes explicitly, defining them rigorously up to basis transformations
in the orthogonal-complement space at each level. The constructive proof of
Eq. \eqref{genspan} is thus complete.

The algorithm is not efficient, because it finds all the states, while the
number of trivial states rises quickly even as the shapes die out. E.g.\ for
$N=3$ particles in $d=3$ dimensions, the total number of shapes is $3!^2=36$.
The shape polynomial reads
\be
P_3(3,q)=q^9 + 3q^7 + 7q^6 + 6q^5 + 6q^4 + 10q^3 + 3q^2
\label{fermN3d3}
\ee
(note the triply degenerate ground state), so there is a single shape in the
seventh excited level (coefficient of $q^9$). But the degeneracy of the
seventh level is $3838$, so the algorithm spends most of its time finding the
span of the $3837$ trivial states, in order to extract the last single shape.

The $d$-th power appearing in the factorization \eqref{harmansatz}, as
reflected by Eq.~\eqref{generating} in the above construction, allows a
refinement of the expression \eqref{genspan} in general. Namely, the terms
$\Phi_i$ can always be written, e.g.\ in three dimensions,
\be
\Phi_i=\sum_{jkl}c^i_{jkl}\Phi^x_{ij}\Phi^y_{ik}\Phi^z_{il},
\ee
where the $\Phi^{x,y,z}$ are Euler-boson monomials, each corresponding to a
particular direction in space (Cartesian axis). In other words, the $\Phi_i$
are superpositions of terms \emph{independently} symmetric in the $N$
variables (coordinate components) along each of the $d$ directions in space.
This form is far from the most general one in all $Nd$ coordinates, symmetric
upon exchange of any two particle indices. [For example, terms like
$t_1u_2+t_2u_1$ cannot appear alone, but only embedded in factored expressions
like $(t_1+t_2)(u_1+u_2)$.] It is interesting that such a strong restriction
on the coefficients $\Phi_i$ still generates the whole $N$-fermion Hilbert
space. Physically, it means that the shapes are the only ``genuinely''
$d$-dimensional states; all excitations of the shapes may be reached as if the
directions in space were different colors.

On a lesser note, the trivial states generated by the algorithm are not always
orthogonal, because multiplication of various shapes with different Euler
bosons can generate the same monomials. Experience with standard quantum
chemical calculations~\cite{Reine12} suggests that little would be gained by
orthogonalizing these vectors explicitly, especially because the overlap
matrices among the trivial states are quite sparse. There is a physical
interpretation both of the overlaps, and of the sparseness. The overlap
indicates the possibility that exciting some shape with an Euler boson, and
then shedding a \emph{different} Euler boson, will give another shape. Such
reconfiguration by excitation is observed sometimes, but cannot be too easy if
the shapes are robust, hence the sparseness.

\section{Examples}

\subsection{The case $d=2$ and $N=3$}

The present example serves as an illustration of the algorithm, and of the
inverse mapping which recovers a concrete realization of the shapes from the
abstract representation. It is the simplest non-trivial multidimensional
case. The partition function \eqref{harmansatz} is
\begin{multline}
(q^2+4q^3+q^4)\left[\left(\frac{1}{1-q}\right)\left(\frac{1}{1-q^2}\right)
\left(\frac{1}{1-q^3}\right)\right]^2\\
=
q^2+(1\cdot 2+4)q^3+(1\cdot 5+4\cdot 2+1)q^4+\ldots,
\label{example}
\end{multline}
where $q^2+4q^3+q^4=P_2(3,q)$ is the shape polynomial, predicting
six shapes, one of which is the ground state, Eq. \eqref{g0}.
The first-excited manifold is spanned by six Slater determinants:
\begin{align}
&\left|t_1^2, t_2, 1\right|\equiv g_{11},\quad
\left|t_1u_1, t_2, 1\right|\equiv g_{12},\nonumber\\
&\left|u_1^2, t_2, 1\right|\equiv g_{13},\quad
\left|t_1^2, u_2, 1\right|\equiv g_{14},\\
&\left|t_1u_1, u_2, 1\right|\equiv g_{15},\quad
\left|u_1^2, u_2, 1\right|\equiv g_{16}.\nonumber
\end{align}
The deflation algorithm gives the Euler-boson states at first level:
\begin{align}
&e_1(t)g_0=(t_1+t_2+t_3)g_0=-g_{12}+g_{14},\nonumber\\
&e_1(u)g_0=(u_1+u_2+u_3)g_0=-g_{13}+g_{15},
\label{eulbos1}
\end{align}
where $g_0$ is the ground state \eqref{g0}. The four states orthogonal to them
are the shapes predicted by the term $4q^3$ in $P_2(3,q)$:
\begin{align}
&S_{11}=g_{11},\quad S_{12}=g_{12}+g_{14},\nonumber\\
&S_{13}=g_{13}+g_{15},\quad S_{14}=g_{16}.
\label{fourshapes}
\end{align}

At the second level, spanned by $14$ Slater determinants, the partition
function breaks down the multiplicity as $14=1\cdot 5+4\cdot 2+1$, which
amounts to: (a) the ground-state $g_0$ multiplied by any of $e_1(t)^2$,
$e_1(u)^2$, $e_1(t)e_1(u)$, $e_2(t)$, or $e_2(u)$; (b) any of the four
shapes \eqref{fourshapes} at first level, multiplied by either $e_1(t)$ or
$e_1(u)$; (c) finally the last shape, orthogonal to the $13$ trivial states
just listed. It is
\begin{multline}
S_{2}=|t_1u_1^2, t_2, 1| - |t_1^2u_1, u_2, 1| \\
+ |t_1^2, u_2^2, 1| - |t_1u_1,t_2, u_3|.
\label{lastshape}
\end{multline}

To visualize these states in real space, one must map the abstract
(node-counting) representation back to some concrete realization. A standard
model for electrons in a quantum dot is to place them in a harmonic oscillator
potential~\cite{Warburton98,Kalliakos09}. For the oscillator potential, the
required inverse of the mapping \eqref{map} is
\be
t_i^k\to\phi_k(x_i)=H_k(x_i)e^{-x_i^2/2},
\label{mapping}
\ee
and similarly for the other directions, with $H_k$ the Hermite polynomial.
This mapping operates uniquely only on monomials like $t^ku^m$, because
$t^kt^m=t^{k+m}$ does not imply $\phi_k(x)\phi_m(x)=\phi_{k+m}(x)$. Hence it
should be applied to factored expressions like $\Phi\Psi$ only after
expanding them in the abstract representation first.

\begin{figure}
\includegraphics[width=4cm]{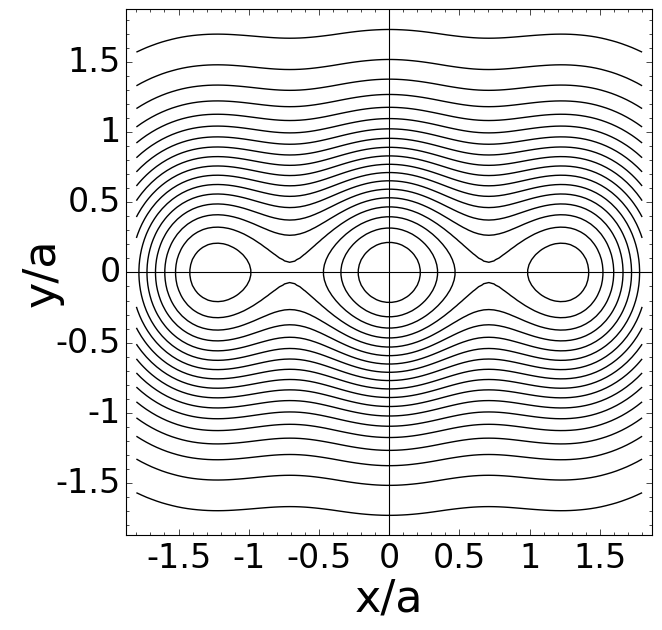}
\includegraphics[width=4cm]{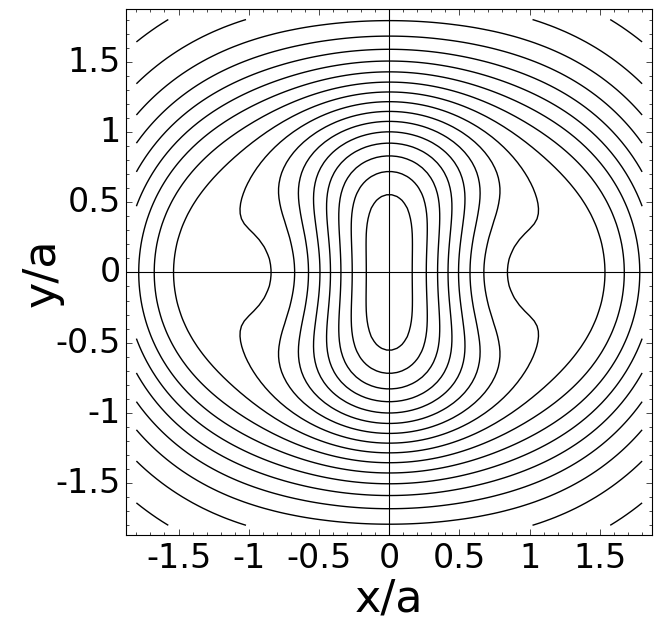}
\includegraphics[width=4cm]{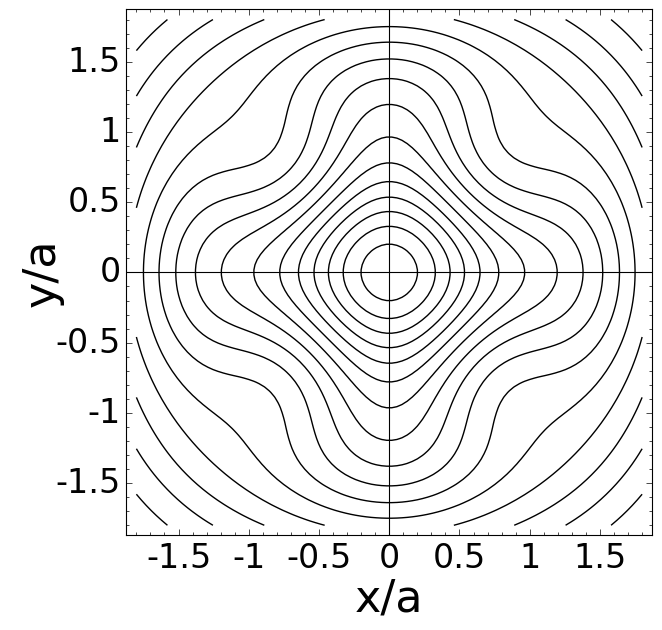}
\caption{Single-particle densities for $S_{11}$, $S_{12}$,
and $S_2$ in the oscillator potential. The coordinate scale is
the oscillator length $a=\sqrt{\hbar/m\omega}$.}.
\label{figrho1}
\end{figure}
The normalized single-particle densities corresponding to $S_{11}$, $S_{12}$,
and $S_2$ are shown in Fig.~\ref{figrho1}. Note that $S_{11}$ is just the
ground state of the one-dimensional system, appearing as a first-excited state
in two dimensions. $S_{13}$ and $S_{14}$ are rotated by $90^\circ$ with
respect to $S_{11}$ and $S_{12}$, so there are only four ``essentially''
different shapes, not six, including the ground state. Obviously, this
redundancy is related to the invariance under relabeling of the axes.

\begin{figure}
\begin{center}
\includegraphics[width=4cm]{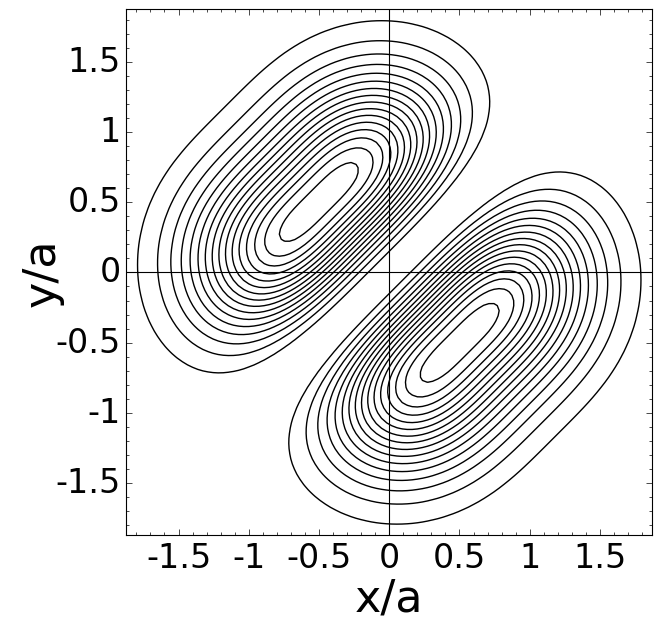}
\includegraphics[width=4cm]{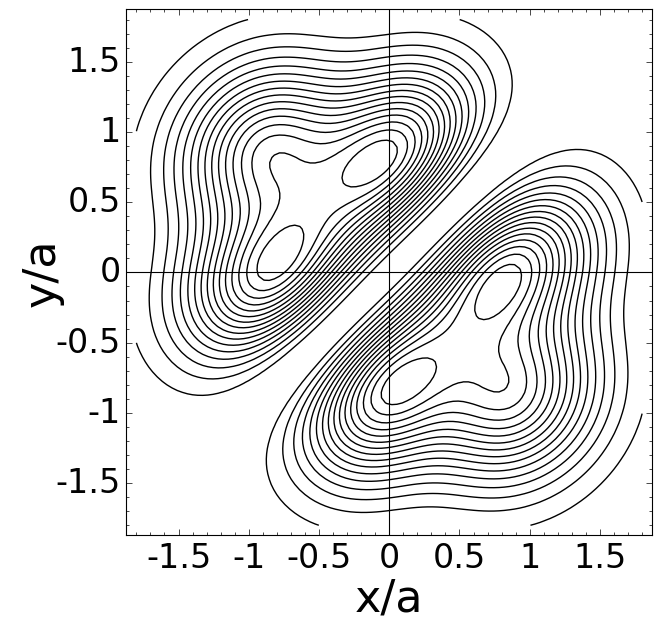}
\end{center}
\caption{Two-particle densities for $e_1(t)g_0$ and $S_{12}$
in the oscillator potential, along the cut $\vec{x}_1=(x,x)$
and $\vec{x}_2=(y,y)$. The scale is the same as in Fig.~\ref{figrho1}.}
\label{figrho2}
\end{figure}
Notably, the shape $S_{12}=g_{12}+g_{14}$ and the trivial state
$e_1(t)g_0=-g_{12}+g_{14}$ have the same single-particle density matrix,
because the Slater determinants $g_{12}$ and $g_{14}$ differ in two orbitals,
so the cross-terms $g_{12}\cdot g_{14}$ vanish when integrated in all but one
variable. This means they are part of the same manifold of wave functions over
which the density functional is determined by minimization in the
Hohenberg-Kohn~\cite{Hohenberg64} approach, for a given density $n$:
\be
F[n]=\min_{\{\Psi: \rho[\Psi]=n\}}
\left<\Psi\left|\hat{H}_0+\hat{V}_{ee}\right|\Psi\right>.
\label{hkohn}
\ee
They differ in the correlation (two-particle) density matrix, as shown in
Fig.~\ref{figrho2}.

The whole discussion above could have been carried out equally well for
electrons in a box, with the mapping
\be
t_i^k\to \cos kx_i,\quad u_i^k\to \cos ky_i
\ee
for open boundary conditions, replacing $\cos kx,y$ with $\sin(k+1)x,y$ for
closed boundary conditions. In any realization, the six shapes span the whole
space of antisymmetric three-body states in two dimensions, using only
symmetric-function coefficients.

\subsection{Coulomb interaction}

For the repulsive Coulomb interaction between fermions in a harmonic well,
small numerical experiments in $d=2$ and $d=3$ invariably favor the shapes as
giving a smaller value of the Coulomb repulsion
$\left<\Psi\right|\hat{V}_{ee}\left|\Psi\right>$, over the trivial basis
states of the form $\Psi_T=\Phi\Psi_1$, where $\Psi_1$ is some shape, excited
by a symmetric term $\Phi\neq 1$, taken nodeless (otherwise anything can be
construed). Plausibly, $\Phi$ seems to act as a coherent amplification for the
final value of the integral, so that $\Phi=1$ gives a smaller integral
overall, which is preferable when the force is repulsive.

Typically, one observes in the numerical experiments that the diagonal Coulomb
matrix elements separate a shape clearly from the multiplet of trivial states,
spanned by the same Slater determinants. E.g., the state \eqref{lastshape},
spanned by four vectors, is separated from the remaining triplet. The
off-diagonal elements also show the expected pattern, in that they are much
smaller among different shapes, than within such multiplets. In other words,
it is much more difficult for the Coulomb force to change a shape, than to
relax a shape over its related multiplet of trivial states.

The observed effects of the Coulomb force conform to the idea that excited
states are organized into bands, such that the lowest state in each band is
dominated by a single shape. Such excitation patterns are ubiquitous in finite
systems, including nuclei, molecules, and quantum dots, where the lowest state
in each band is sometimes called the band-head. Shapes are natural candidates
for the band-head states, because, as noted already, they are the only
genuinely $d$-dimensional states.

\subsection{Trial wave functions}

Truncations of the method which builds the whole Hilbert space give rise to
specific families of trial wave functions. For example, take two particles in
three dimensions. The shape polynomial is $3q+q^3$, and the four shapes are
\begin{align}
&\Psi_1=t_1-t_2,\quad \Psi_2=u_1-u_2,\nonumber\\
&\Psi_3=v_1-v_2,\quad \Psi_4=\Psi_1\Psi_2\Psi_3.
\label{2p3d}
\end{align}
Contrary to the intuitive idea of shapes in the Introduction, $\Psi_4$ does
factor over the space dimensions. Such an ``accidentally'' factored
term must appear whenever it, too, gives a possible way to write an
antisymmetric function. Here it is the only higher shape, showing that the
number of ways an antisymmetric wave function can be constructed is quite
restricted for two particles.

A general two-body wave function can be written by combining the $\Psi_i$ with
symmetric-polynomial coefficients:
\be
\sum_{i=1}^4\Phi_i(t,u,v)\Psi_i.
\label{exp2p3d}
\ee
Trial wave functions are obtained by restricting the polynomials in various
ways. For example, $\Phi_i(t,u,v)=c_{i0}+c_{i1}e_1(t)+c_{i2}e_1(u)+
c_{i3}e_1(v)$. The approach provides a qualitative language to describe the
trial functions. Thus, whether $e_2$ or $e_1^2$ is more important in
second order is a question with physical meaning, because, as noted before,
$e_2$ is a gas-like excitation, while $e_1^2$ is liquid-like.

The trial wave functions use a relatively small number of Slater determinants,
because the shapes themselves are quite sparing in this sense. E.g.\ $S_2$
in Eq. \eqref{lastshape} uses only four vectors of the available fourteen.
Similarly, the highest shape for $N=3$ in $d=3$ is spanned by $36$ Slater
determinants out of possible $3838$. Finding that particular combination is
quite beyond trial and error.

\subsection{Simulations}

The critical issue is to locate the nodes of many-body functions in the
$Nd$-dimensional configuration space~\cite{Ceperley91}, otherwise the
well-known fermion sign problem appears~\cite{Loh90}. These nodes are unknown,
so that simulations use a guiding function whose nodes are supposed to be near
those of the exact solution. The shapes provide a finite and complete
antisymmetric-function basis for guiding functions in simulations.

The model of spin-polarized electrons confined to a sphere in $d=3$ is of
contemporary interest as a test-bed for theory and simulations~\cite{Loos15}.
Here it means mapping $t,u,v$ to $x,y,z$ and interpreting the latter
(Cartesian) coordinates in terms of polar and azimuthal angles on the unit
sphere. For two particles, the shapes \eqref{2p3d} define the nodal surfaces
\be
x_1=x_2,\quad y_1=y_2, \quad z_1=z_2.
\label{realcond}
\ee
In the space of particle $1$, these are three circles which cut off a cap of
the sphere at the coordinates $x_2,y_2,z_2$, respectively, of particle $2$. If
the Hamiltonian is invariant under coordinate permutations, then
$\left<\Psi_i\right|H\left|\Psi_j\right>=0$ for $i\neq j$, $i,j=1,2,3$, so one
can choose any one of them for the guiding function without loss of
generality, say $\Psi_3=z_1-z_2$. Then the most general ground-state wave
function \eqref{exp2p3d} up to relabeling the axes is
\be
\left[\Phi_1+(x_1-x_2)(y_1-y_2)\Phi_2\right](z_1-z_2)\equiv
(z_1-z_2)\widetilde{\Phi},
\label{gensphere}
\ee
including $\Psi_4$. It follows that the interacting ground state has the same
nodes $z_1=z_2$ as the non-interacting ground state
$\widetilde{\Phi}=\mbox{\emph{const.}}$,
assuming~\cite{Ceperley91,Bajdich05,Loos15} that
$\widetilde{\Phi}\neq\mbox{\emph{const.}}$ does not introduce new nodes. This
result was recently derived as a theorem for this particular
model~\cite{Loos15}, while the above reasoning is model-independent,
based on the limited number of possible shapes, as listed in Eq. \eqref{2p3d}.

By the same reasoning, a similar result as \eqref{gensphere} can be obtained
for the oscillator potential, with the mapping \eqref{mapping}. Then the
interesting question arises, whether simulating the oscillator in real or
complex (Bargmann) space is more convenient, given that complexification
doubles the number of real variables. A simulation keeping $z_1<z_2$ should
converge to a form like \eqref{gensphere} for a nodeless $\widetilde{\Phi}$,
however the natural eigenfunctions of the problem are still the Hermite
functions, which can be recovered only by multiplying out the original
abstract expression:
\be
(v_1-v_2)\widetilde{\Phi}\to
\sum_{\vec{n}} c_{\vec{n}} \Phi_{\vec{n}}(\vec{R}),
\ee
where $\Phi_{\vec{n}}(\vec{R})$ contains Hermite functions. The node
$v_1=v_2$ which was explicitly controlled in the Bargmann representation is
now hidden under cancellations of oscillating functions. Real-space
representations generically have the problem that nodes of the constituent
one-body wave functions, required by orthogonality, interfere with the
analysis of nodes of the $N$-body function,
which are completely different objects~\cite{Ceperley91}.

Similar issues arise for standing waves $\cos kx$, which may be avoided by
the travelling-wave complexification $e^{ikx}$. As of this writing, it seems
that the advantages of having $\phi_k\phi_m=\phi_{k+m}$ outweigh any
disadvantage of complexification. Further considerations along these lines are
beyond the scope of this article.

\section{Space dimension and boson-fermion correspondence}

The factorization \eqref{Zdfact} is the same for identical bosons. The only
difference in the recursion for the shape polynomial \eqref{recurPd} is that
the alternating sign $(-1)^{k+1}$ does not appear, and Slater determinants
have to be replaced by permanents (i.e.\ lose the alternating sign) in the
general algorithm. The Euler bosons remain formally the same elementary
symmetric function monomials. Indistinguishable (original) bosons are replaced
by distinguishable Euler bosons and shapes, in close parallel to the fermion
case. This correspondence explains where have ``gone'' all the most general
symmetric functions, alluded to in Sect.~\ref{genalg} above. They span the
space of identical bosons, which is however also finitely generated, with
coefficients (Euler bosons) as restricted as the ones for fermions. In other
words, just as a finite number of antisymmetric $N$-body functions is
sufficient to generate them all, so can all symmetric functions be generated
from a finite number of genuinely $d$-dimensional bosonic shapes. These
symmetric shapes are the only ``real'' difference between bosons and fermions.

An interesting distinction appears between spaces of odd and even dimension.
In even dimensions, shape polynomials are always symmetric. This can be
understood by replacing $q\to 1/q$ in the recursion \eqref{recurPd}, which
reverses the polynomial. Clearly the net effect on the recursion is that the
coefficient $C_k^N(q)$ gains an extra sign of $(-1)^{k+1}$. Because it is
raised to the $d$-th power, this extra sign vanishes in even dimensions, so
the recursion for the polynomial and for the reversed polynomial is the same.
Therefore the polynomial must be symmetric, in both bosonic and fermionic
cases. Such is $q^2+4q^3+q^4$ in Eq. \eqref{example}.

The odd-dimensional case is more interesting. Now the sign change $(-1)^{k+1}$
cancels the $(-1)^{k+1}$ in the recursion for the fermionic case, and
introduces it in the bosonic case: polynomial reversal changes the bosonic
recursion into the fermionic one, and vice versa. This means that the
coefficient lists in the shape polynomials for bosons and for fermions are
``mirror images'' of each other. For example, the shape polynomial for $N=3$
bosons in $d=3$ dimensions is
\be
B_3(3,q)=1 + 3q^2 + 7q^3 + 6q^4 + 6q^5 + 10q^6 + 3q^7,
\label{bosN3d3}
\ee
to be compared with $P_3(3,q)$ in Eq.~\eqref{fermN3d3}.

In odd dimensions, physical inferences can be made between the bosonic case
and the fermionic one. For example, every bosonic polynomial $B_d(N,q)$ begins
with a coefficient of unity, because the bosonic ground state cannot be
degenerate --- but this statement means that the \emph{highest} shape in the
fermionic polynomial is always non-degenerate. Furthermore, the first excited
state for bosons contains no shapes, because its degeneracy is always $d$, and
there are also $d$ Euler bosons $e_1$, as mentioned before. Therefore, there
will be no shapes at the \emph{second-highest} level for fermions --- the
absence of the term $q^8$ in Eq. \eqref{fermN3d3} mirrors the absence of $q^1$
in Eq. \eqref{bosN3d3}. For a simple example, $\Psi_4$ in Eq.~\eqref{2p3d} is
a second-excited state, while all nine first-excited states are of the form
$(a_1+a_2)(b_1-b_2)$ with $a,b=t,u,v$.

\section{Discussion}

The main result of this work is a fundamental insight into the structure of
$N$-fermion Hilbert space: a finite number of antisymmetric functions generate
all antisymmetric functions, with symmetric-function coefficients. This
property of being \emph{finitely generated} does not depend on any particular
realization of the Hilbert space~\cite{Derksen02}. It has been made explicit
here with the convenient choice of one-body functions separable in Cartesian
coordinates. The independence of the main result on such technicalities was
demonstrated in the formula~\eqref{asymptot}, which needed only the structural
formula~\eqref{recz} to count the shapes directly from the
\emph{ansatz}~\eqref{Zdfact}.

Mathematically, Eq. \eqref{harmansatz} is a Poincar\'e (a.k.a.\ Hilbert)
series~\cite{Stanley78}, which counts the dimensions of the vector (Hilbert)
spaces of a given grade, which may be visualized as the degeneracy of the
corresponding oscillator level. The particular form of the series indicates
how these spaces may be generated algebraically by combining certain invariant
polynomials, called Euler bosons and shapes here. In the standard language of
invariant theory~\cite{Derksen02}, the Euler bosons are \emph{primary}, and
the shapes \emph{secondary} invariants. This identification
follows~\cite{Derksen02} from the most general form of the wave function,
e.g.\ for $d=3$:
\be
\sum_{i=1}^{N!^2} \Phi_i\Psi_i,
\quad
\Phi_i=\sum_{jkl}c^i_{jkl}\Phi^x_{ij}\Phi^y_{ik}\Phi^z_{il},
\label{exppsi}
\ee
where the $\Phi^{x,y,z}$ are monomials of Euler bosons in the three
directions, while the $\Psi_i$ are all the $N!^2$ shapes of $N$ particles in
three dimensions. The invariants $\Psi_i$ are antisymmetric polynomials over
$\mathbb{Z}$ which finitely generate the Hilbert space of $N$ identical
fermions as a graded algebra, with coefficients $\Phi_i$ from the ring of
polynomials over $\mathbb{C}$, independently symmetric in each of $d$ sets of
$N$ variables. The grading is by degree of the polynomials, which is just the
energy in the oscillator case. Remarkably, but not unexpectedly, the main
result \eqref{exppsi} is equally valid for bosons and for fermions, with
symmetry in place of antisymmetry, and permanents replacing determinants in
the constitutive expressions.

Antisymmetric polynomials in one dimension can always be studied by proxy
symmetric polynomials: Slater determinants in formal powers and Schur
functions differ by a fixed factor, the Vandermonde
determinant~\cite{Stanley99}. The present work shows that when $d>1$
antisymmetry gives rise to qualitatively new polynomial invariants, the
shapes. They are a different generalization of the Vandermonde determinant
than the obvious one, which is just an excited one-dimensional state. This
mathematical generalization has a direct physical meaning as the
generalization of the Slater-determinant form for the ground state to more
than one dimension. The appearance of additional antisymmetric invariants ---
the shapes --- is a consequence of the \emph{weakening} of the Pauli principle
when $d>1$, because it requires antisymmetry only with respect to interchange
of \emph{vector} coordinates, i.e.\ simultaneous interchange of $d$-plets of
variables refering to different particles, as opposed to the interchange of
any two variables, which is the case in one dimension.

Particles with different spin projections are distinguishable, so their wave
functions can be obtained by a simple direct product of the spaces discussed
here. Notably, the shape space is not closed with respect to spin. For $2N$
spin-up fermions, there are $(2N)!^{d-1}$ shapes, while for $N$ fermions of
spin up and $N$ of spin down, there are only $N!^{2(d-1)}$ shapes, a much
smaller number in general. Raising the total spin projection, which makes more
particles indistinguishable, increases the choice of shapes, i.e.\ orbital
states with enhanced collectivity. This observation fits well with Hund's
rule~\cite{Yamanaka05}: spin-polarized states are preferred when Coulomb
effects are important.

The direct product of up- and down-spin spaces does not imply that the wave
functions have to be in pure product form, which is known to constrain them
unphysically~\cite{Bressanini12}. One can assume a configuration-interaction
(CI) form, which is the superposition
\be
\sum_ic_i\Psi_{\uparrow i}\Psi_{\downarrow i},
\ee
where the $\Psi_{\sigma i}$ are particular cases of \eqref{exppsi}. A CI form
can describe the topology of the exact nodal surface~\cite{Bressanini12}. By
mapping all wave functions onto symbolic polynomials, the algebraic approach
puts the discussion of nodal-surface topologies directly into the purview of
algebraic geometry, one of whose traditional concerns are the zeros of
multivariate polynomials~\cite{MilneAG,Dieudonne72}. On the other hand, there
is always an underlying differentiable manifold, spanned by the original
one-body wave functions.  For the harmonic oscillator, the Bargmann transform
even allows a direct reinterpretation of the same polynomials as analytic
functions in complex (Bargmann) space. Physical intuition suggests that
possible nodal-surface topologies should not depend qualitatively on the
confining potential, as long as one can be adiabatically transformed into
another. Hence the harmonic-oscillator setting is already quite general, as
far as the topology of nodal surfaces is concerned.

Restoration of rotational invariance similarly proceeds by superposition. As
already noted, such basis issues cannot impinge on the underlying property of
Hilbert space, that it is finitely generated. However, a large part of
practical invariant theory~\cite{Derksen02} is to find optimal sets of
generators for particular applications, and the flexibility of the algebraic
structure in the choice of generators bodes well for future physics
applications. In the present work, an explicit realization of generators
organized by grade has been given. They can either be ``post-processed'' into
rotationally invariant states, or perhaps a completely different algorithm may
be found which produces rotationally invariant shapes natively. Generally,
restoration of symmetries broken by the shape-generating algorithm is needed
whenever they are not broken by the physical ground state.

The existence of shapes provides an unexpected perspective on the fermion sign
problem~\cite{Loh90}. Given that the Euler-boson wave functions are symmetric,
the fermion sign problem appears only because there exists more than one
shape. Conversely, if a problem could be described by the excitations of a
single shape, the whole physical space of the system could be described in the
Euler boson language, avoiding the sign problem. One can envisage
\emph{imposing} such a scenario in a Kohn-Sham-like approach, choosing a
particular shape by qualitative argument, and making it give the correct
binding energy with a self-consistently derived single-particle basis. Such a
program is conceptually similar to a fixed-node
approach~\cite{Bressanini12,Filinov14}, except that some movement of the nodes
is still allowed, due to optimization in the Euler-boson sector.

The finite number of shapes brings variation and simulation closer together
than is usually understood. The fact that $e^{-\tau H}$ is a general projector
on the exact ground state becomes relative when generality is a finite range
of possibilities, listed in advance. It is then a matter of expediency rather
than principle to replace the universal projector $e^{-\tau H}$ with a
specific projector in a given simulation. An explicit choice of ground-state
projector turns a simulation into variational optimization.

From a practical point of view, the factorial rise in the number of shapes is
somewhat unfortunate. However, problems involving strong correlations are
usually local in nature, i.e.\ involve only a small number of electrons. Even
in solid-state physics, this case is common, as attested by the remarkable
popularity of locally based approaches, from finite-system studies to
dynamical mean-field theory~\cite{Georges96}. Taking
$N_\uparrow=N_\downarrow=4$ as a modestly ambitious limit of practicality for
$d=3$ --- meaning that 255 of the 576 shapes for this case have been generated
on the author's laptop, while the rest would require additional optimization
and/or a bigger computer --- problems with up to eight unpolarized electrons
are within reach, which is competitive as of this writing. In two dimensions,
the situation is naturally better.

The expression \eqref{exppsi} collects the two main results of this work.
First, there is a \emph{finite} number of shapes in which any wave function
can be expanded. In physical terms, there is a finite number of possible
$N$-body vacua. Second, the polynomial coefficients in this expansion, or
excitations of the vacua, are 1D-bosonic, i.e.\ symmetric in the $N$ space
coordinates on each of the $d$ axes separately.

To conclude, the notion of the $N$-body vacuum in $d$ space dimensions has
been given a precise and general algebraic meaning for fixed $N$ and $d$. An
algorithm to construct all possible vacua was presented, and it was shown that
they finitely generate the full Hilbert space of $N$ identical particles. It
is hoped that these insights will lead to advances in practical calculation,
at least for values of $N$ similar to those encountered in contemporary work.

\acknowledgments

I thank D.~Svrtan for his help and interest in this work. Conversations with
O.~S.~Bari\v si\'c, I.~Batisti\'c, and M.~Primc are gratefully acknowledged.
Thanks are also due to J.~Cios\l owski, P.-F. Loos, J.~Tahir-Kheli,
H.~Van\v cik and J.~Zaanen for reading and commenting upon the manuscript.
This work was supported by the Croatian Ministry of Science grant
119-1191458-0512 and by the University of Zagreb grant 202759.

\appendix
\begin{widetext}
\section{Notes on Schur functions~\cite{Stanley99}}

The classic definition of Schur functions is a ratio of two determinants. The
denominator is the Vandermonde determinant in some indeterminates $z_i$,
\be
\Delta(z_1,z_2,\ldots,z_N)\equiv\left|
\begin{matrix}
z_1^{N-1}&z_2^{N-1}&\cdots&z_N^{N-1}\\
z_1^{N-2}&z_2^{N-2}&\cdots&z_N^{N-2}\\
\vdots&\vdots&&\vdots\\
z_1&z_2&\cdots&z_N\\
1&1&\cdots&1
\end{matrix}
\right|=\prod_{1\le i<j\le N}(z_i-z_j),
\ee
which physicists would call a ground-state Slater determinant. The numerator
is a similar determinant with some higher powers of the $z_i$ --- an excited
state in physicists' terms, while mathematicians sometimes call it a
generalized Vandermonde determinant. If $\lambda=(\lambda_1,\ldots,\lambda_N)$
is a non-increasing sequence of natural numbers or zeros (a \emph{partition}
of the number $|\lambda|\equiv\lambda_1+\ldots+\lambda_N$ into at most $N$
parts), then the Schur function $s_\lambda$ is defined by
\be
s_\lambda\equiv\frac{1}{\Delta(z_1,z_2,\ldots,z_N)}\left|
\begin{matrix}
z_1^{N-1+\lambda_1}&z_2^{N-1+\lambda_1}&\cdots&z_N^{N-1+\lambda_1}\\
z_1^{N-2+\lambda_2}&z_2^{N-2+\lambda_2}&\cdots&z_N^{N-2+\lambda_2}\\
\vdots&\vdots&&\vdots\\
z_1^{\lambda_N}&z_2^{\lambda_N}&\cdots&z_N^{\lambda_N}
\end{matrix}
\right|.
\ee
The divisibility of the numerator by the denominator may be inferred from the
fact that both vanish when any two $z_i=z_j$. The result of the division is
given by a combinatorial interpretation of $s_\lambda$. Take a Young tableau
of shape $\lambda$ and fill it with natural numbers not greater than $N$,
increasing along columns and nondecreasing along rows. Call $n_k\ge 0$ the
number of times the number $k$ appears in the tableau. A \emph{type}
$T(\lambda)$ is just a particular filling so obtained, for a given shape
$\lambda$; then
\be
s_\lambda=\sum_{T(\lambda)}z_1^{n_1}z_2^{n_2}\cdots z_N^{n_N},
\ee
where the sum is over all possible types. Thus coefficients in Schur functions
must be natural (counting) numbers. Operationally, this formula is much
simpler than the determinantal one. For example,
\be
s_1=z_1+z_2+\ldots+z_N,
\ee
because a single box can be filled with the numbers $1,2,\ldots,N$ only one at
a time. On the other hand, if all the $\lambda_i=1$, this corresponds to a
vertical strip of height $N$, which can be filled in only one way,
\be
s_{1^N}=z_1z_2\cdots z_N.
\ee
The elementary symmetric functions $e_k$ similarly correspond to vertical
strips of height $k$: they are the Schur functions of the partition
$\lambda=(1\ldots 1)_{\mbox{\scriptsize k times}}=1^k$. E.g.\ for $k=2$ in
$N=3$ variables, $s_{11}=s_{1^2}=z_1z_2+z_1z_3+z_2z_3$.

\section{Matrix element of the Coulomb force}

In quantum chemical calculations, one typically uses matrix elements between
non-orthogonal Hermite Gaussian functions~\cite{Zivkovic68}, which are best
calculated recursively~\cite{McMurchie78}. I was not able to locate the
corresponding closed expression for orthogonal Hermite functions in the
literature, so I give it here, without pretense to originality.

Let
\be
\left[\vec{n}\vec{n}'\right|V_C\left|\vec{m}\vec{m}'\right]=
\int d\vec{R}\,d\vec{R}'\,
\Phi_{\vec{n}}^*(\vec{R})\Phi_{\vec{n}'}^*(\vec{R}')
\frac{1}{|\vec{R}-\vec{R}'|}
\Phi_{\vec{m}}(\vec{R})\Phi_{\vec{m}'}(\vec{R}')
\ee
be the two-body matrix element between products of unnormalized
Hermite functions,
\be
\Phi_{\vec{n}}(\vec{R})\equiv
\phi_{n_1}(R_1)\cdots\phi_{n_d}(R_d),\quad \phi_n(x)=H_n(x)e^{-x^2/2},
\label{defPhi}
\ee
where $H_n$ is the Hermite polynomial. Using the standard trick~\cite{Reine12}
\be
\frac{1}{|r|}=\frac{1}{\sqrt{\pi}}\int_{-\infty}^{+\infty} e^{-r^2w^2}dw,
\ee
one finds, for dimensions $d>1$:
\be
\left[\vec{n}\vec{n}'\right|V_C\left|\vec{m}\vec{m}'\right]=
\pi^d\sqrt{\frac{2}{\pi}}\int_0^1 dw\,(1-w^2)^{(d-3)/2}\prod_{i=1}^d
\sum_{k_i=0}^{n_i+m_i}\sum_{k_i'=0}^{n_i'+m_i'}
a^{n_im_i}_{k_i}a^{n_i'm_i'}_{k_i'}
(-1)^{k_i}H_{k_i+k_i'}(0)
\left(\frac{w}{\sqrt{2}}\right)^{k_i+k_i'},
\label{coulomb}
\ee
where
\be
a^{nm}_{k}=\frac{2^{(n+m-k)/2}n!m!}
{\left(\frac{m+n-k}{2}\right)!\left(\frac{k+n-m}{2}\right)!
\left(\frac{k+m-n}{2}\right)!}
\ee
for $n+m+k$ even and non-negative factorials in the denominator, zero
otherwise. The Hermite polynomials $H_{k+k'}(0)$, evaluated at zero, are zero
for $k+k'$ odd, and
\be
(-1)^k H_{k+k'}(0)=
(-1)^{(k-k')/2}\frac{(k+k')!}{\left(\frac{k+k'}{2}\right)!}
\ee
for $k+k'$ even. Finally, when the product in Eq. \eqref{coulomb} is expanded,
the integrals over $w$ give the beta function in place of Boys'
function~\cite{Reine12}:
\be
I_d(l)=\int_0^1(1-w^2)^{(d-3)/2}w^l\,dw
=B\left(\frac{l+1}{2},\frac{d-1}{2}\right)=
\frac{\Gamma\left(\frac{l+1}{2}\right)\Gamma\left(\frac{d-1}{2}\right)}
{2\Gamma\left(\frac{l+d}{2}\right)}.
\ee
In particular,
\be
I_2(l)=\frac{\pi}{2^{l+1}}\binom{l}{l/2}\approx
\sqrt{\frac{\pi}{2l}},\quad
I_3(l)=\frac{1}{l+1},
\ee
noting that $l=\sum_ik_i+k_i'$ is always even.

\end{widetext}


\begin{thebibliography}{31}%
\makeatletter
\providecommand \@ifxundefined [1]{%
 \@ifx{#1\undefined}
}%
\providecommand \@ifnum [1]{%
 \ifnum #1\expandafter \@firstoftwo
 \else \expandafter \@secondoftwo
 \fi
}%
\providecommand \@ifx [1]{%
 \ifx #1\expandafter \@firstoftwo
 \else \expandafter \@secondoftwo
 \fi
}%
\providecommand \natexlab [1]{#1}%
\providecommand \enquote  [1]{``#1''}%
\providecommand \bibnamefont  [1]{#1}%
\providecommand \bibfnamefont [1]{#1}%
\providecommand \citenamefont [1]{#1}%
\providecommand \href@noop [0]{\@secondoftwo}%
\providecommand \href [0]{\begingroup \@sanitize@url \@href}%
\providecommand \@href[1]{\@@startlink{#1}\@@href}%
\providecommand \@@href[1]{\endgroup#1\@@endlink}%
\providecommand \@sanitize@url [0]{\catcode `\\12\catcode `\$12\catcode
  `\&12\catcode `\#12\catcode `\^12\catcode `\_12\catcode `\%12\relax}%
\providecommand \@@startlink[1]{}%
\providecommand \@@endlink[0]{}%
\providecommand \url  [0]{\begingroup\@sanitize@url \@url }%
\providecommand \@url [1]{\endgroup\@href {#1}{\urlprefix }}%
\providecommand \urlprefix  [0]{URL }%
\providecommand \Eprint [0]{\href }%
\providecommand \doibase [0]{http://dx.doi.org/}%
\providecommand \selectlanguage [0]{\@gobble}%
\providecommand \bibinfo  [0]{\@secondoftwo}%
\providecommand \bibfield  [0]{\@secondoftwo}%
\providecommand \translation [1]{[#1]}%
\providecommand \BibitemOpen [0]{}%
\providecommand \bibitemStop [0]{}%
\providecommand \bibitemNoStop [0]{.\EOS\space}%
\providecommand \EOS [0]{\spacefactor3000\relax}%
\providecommand \BibitemShut  [1]{\csname bibitem#1\endcsname}%
\let\auto@bib@innerbib\@empty
\bibitem [{\citenamefont {Kohn}\ and\ \citenamefont {Sham}(1965)}]{Kohn65}%
  \BibitemOpen
  \bibfield  {author} {\bibinfo {author} {\bibfnamefont {W.}~\bibnamefont
  {Kohn}}\ and\ \bibinfo {author} {\bibfnamefont {L.~J.}\ \bibnamefont
  {Sham}},\ }\href {\doibase 10.1103/PhysRev.140.A1133} {\bibfield  {journal}
  {\bibinfo  {journal} {Phys. Rev.}\ }\textbf {\bibinfo {volume} {140}},\
  \bibinfo {pages} {A1133} (\bibinfo {year} {1965})}\BibitemShut {NoStop}%
\bibitem [{\citenamefont {Dirac}(1926)}]{Dirac26}%
  \BibitemOpen
  \bibfield  {author} {\bibinfo {author} {\bibfnamefont {P.~A.~M.}\
  \bibnamefont {Dirac}},\ }\href {\doibase 10.1098/rspa.1926.0133} {\bibfield
  {journal} {\bibinfo  {journal} {Proceedings of the Royal Society of London A:
  Mathematical, Physical and Engineering Sciences}\ }\textbf {\bibinfo {volume}
  {112}},\ \bibinfo {pages} {661} (\bibinfo {year} {1926})}\BibitemShut
  {NoStop}%
\bibitem [{\citenamefont {Slater}(1929)}]{Slater29}%
  \BibitemOpen
  \bibfield  {author} {\bibinfo {author} {\bibfnamefont {J.~C.}\ \bibnamefont
  {Slater}},\ }\href {\doibase 10.1103/PhysRev.34.1293} {\bibfield  {journal}
  {\bibinfo  {journal} {Phys. Rev.}\ }\textbf {\bibinfo {volume} {34}},\
  \bibinfo {pages} {1293} (\bibinfo {year} {1929})}\BibitemShut {NoStop}%
\bibitem [{\citenamefont {Derksen}\ and\ \citenamefont
  {Kemper}(2002)}]{Derksen02}%
  \BibitemOpen
  \bibfield  {author} {\bibinfo {author} {\bibfnamefont {H.}~\bibnamefont
  {Derksen}}\ and\ \bibinfo {author} {\bibfnamefont {G.}~\bibnamefont
  {Kemper}},\ }\href@noop {} {\emph {\bibinfo {title} {Computational Invariant
  Theory}}},\ Encyclopaedia of Mathematical Sciences\ (\bibinfo  {publisher}
  {Springer Berlin Heidelberg},\ \bibinfo {year} {2002})\BibitemShut {NoStop}%
\bibitem [{\citenamefont {Euler}(1751)}]{Euler51}%
  \BibitemOpen
  \bibfield  {author} {\bibinfo {author} {\bibfnamefont {L.}~\bibnamefont
  {Euler}},\ }\href@noop {} {\bibfield  {journal} {\bibinfo  {journal} {Comm.
  Acad. Petrop.}\ }\textbf {\bibinfo {volume} {13}},\ \bibinfo {pages} {64}
  (\bibinfo {year} {1741--43, 1751})},\ \bibinfo {note} {see p.\ 80, or
  arxiv:0711.3656 (transl.\ by J.~Bell), p.\ 15. Euler gets $q^{N(N+1)/2}$
  because he starts from $1\leq n_1$ in Eq.~\eqref{Z1}.}\BibitemShut {Stop}%
\bibitem [{\citenamefont {King}\ and\ \citenamefont {Welsh}(2006)}]{King06}%
  \BibitemOpen
  \bibfield  {author} {\bibinfo {author} {\bibfnamefont {R.~C.}\ \bibnamefont
  {King}}\ and\ \bibinfo {author} {\bibfnamefont {T.~A.}\ \bibnamefont
  {Welsh}},\ }\href {\doibase 10.1088/1742-6596/30/1/001} {\bibfield  {journal}
  {\bibinfo  {journal} {Journal of Physics: Conference Series}\ }\textbf
  {\bibinfo {volume} {30}},\ \bibinfo {pages} {1} (\bibinfo {year}
  {2006})}\BibitemShut {NoStop}%
\bibitem [{\citenamefont {Warburton}\ \emph {et~al.}(1998)\citenamefont
  {Warburton}, \citenamefont {Miller}, \citenamefont {D\"urr}, \citenamefont
  {B\"odefeld}, \citenamefont {Karrai}, \citenamefont {Kotthaus}, \citenamefont
  {Medeiros-Ribeiro}, \citenamefont {Petroff},\ and\ \citenamefont
  {Huant}}]{Warburton98}%
  \BibitemOpen
  \bibfield  {author} {\bibinfo {author} {\bibfnamefont {R.~J.}\ \bibnamefont
  {Warburton}}, \bibinfo {author} {\bibfnamefont {B.~T.}\ \bibnamefont
  {Miller}}, \bibinfo {author} {\bibfnamefont {C.~S.}\ \bibnamefont {D\"urr}},
  \bibinfo {author} {\bibfnamefont {C.}~\bibnamefont {B\"odefeld}}, \bibinfo
  {author} {\bibfnamefont {K.}~\bibnamefont {Karrai}}, \bibinfo {author}
  {\bibfnamefont {J.~P.}\ \bibnamefont {Kotthaus}}, \bibinfo {author}
  {\bibfnamefont {G.}~\bibnamefont {Medeiros-Ribeiro}}, \bibinfo {author}
  {\bibfnamefont {P.~M.}\ \bibnamefont {Petroff}}, \ and\ \bibinfo {author}
  {\bibfnamefont {S.}~\bibnamefont {Huant}},\ }\href {\doibase
  10.1103/PhysRevB.58.16221} {\bibfield  {journal} {\bibinfo  {journal} {Phys.
  Rev. B}\ }\textbf {\bibinfo {volume} {58}},\ \bibinfo {pages} {16221}
  (\bibinfo {year} {1998})}\BibitemShut {NoStop}%
\bibitem [{\citenamefont {Kalliakos}\ \emph {et~al.}(2009)\citenamefont
  {Kalliakos}, \citenamefont {Rontani}, \citenamefont {Pellegrini},
  \citenamefont {Pinczuk}, \citenamefont {Shinga}, \citenamefont {Garcia},
  \citenamefont {Goldoni}, \citenamefont {Molinari}, \citenamefont {Pfeiffer},\
  and\ \citenamefont {West}}]{Kalliakos09}%
  \BibitemOpen
  \bibfield  {author} {\bibinfo {author} {\bibfnamefont {S.}~\bibnamefont
  {Kalliakos}}, \bibinfo {author} {\bibfnamefont {M.}~\bibnamefont {Rontani}},
  \bibinfo {author} {\bibfnamefont {V.}~\bibnamefont {Pellegrini}}, \bibinfo
  {author} {\bibfnamefont {A.}~\bibnamefont {Pinczuk}}, \bibinfo {author}
  {\bibfnamefont {A.}~\bibnamefont {Shinga}}, \bibinfo {author} {\bibfnamefont
  {C.}~\bibnamefont {Garcia}}, \bibinfo {author} {\bibfnamefont
  {G.}~\bibnamefont {Goldoni}}, \bibinfo {author} {\bibfnamefont
  {E.}~\bibnamefont {Molinari}}, \bibinfo {author} {\bibfnamefont
  {L.}~\bibnamefont {Pfeiffer}}, \ and\ \bibinfo {author} {\bibfnamefont
  {K.}~\bibnamefont {West}},\ }\href {\doibase 10.1016/j.ssc.2009.04.034}
  {\bibfield  {journal} {\bibinfo  {journal} {Solid State Communications}\
  }\textbf {\bibinfo {volume} {149}},\ \bibinfo {pages} {1436} (\bibinfo {year}
  {2009})}\BibitemShut {NoStop}%
\bibitem [{\citenamefont {Miao}\ and\ \citenamefont {Hoffmann}(2014)}]{Miao14}%
  \BibitemOpen
  \bibfield  {author} {\bibinfo {author} {\bibfnamefont {M.-S.}\ \bibnamefont
  {Miao}}\ and\ \bibinfo {author} {\bibfnamefont {R.}~\bibnamefont
  {Hoffmann}},\ }\href {\doibase 10.1021/ar4002922} {\bibfield  {journal}
  {\bibinfo  {journal} {Accounts of Chemical Research}\ }\textbf {\bibinfo
  {volume} {47}},\ \bibinfo {pages} {1311} (\bibinfo {year}
  {2014})}\BibitemShut {NoStop}%
\bibitem [{\citenamefont {Andrews}(1976)}]{Andrews76}%
  \BibitemOpen
  \bibfield  {author} {\bibinfo {author} {\bibfnamefont {G.~E.}\ \bibnamefont
  {Andrews}},\ }\href@noop {} {\emph {\bibinfo {title} {The Theory of
  Partitions}}}\ (\bibinfo  {publisher} {Addison-Wesley},\ \bibinfo {address}
  {New York},\ \bibinfo {year} {1976})\BibitemShut {NoStop}%
\bibitem [{\citenamefont {{Fa\`a di Bruno}}(1881)}]{Bruno81}%
  \BibitemOpen
  \bibfield  {author} {\bibinfo {author} {\bibfnamefont {F.}~\bibnamefont
  {{Fa\`a di Bruno}}},\ }\href@noop {} {\emph {\bibinfo {title} {{Einleitung in
  die Theorie der Bin\"aren Formen}}}}\ (\bibinfo  {publisher} {B. G.
  Teubner},\ \bibinfo {address} {Leipzig},\ \bibinfo {year} {1881})\BibitemShut
  {NoStop}%
\bibitem [{\citenamefont {Ford}(1971)}]{Ford71}%
  \BibitemOpen
  \bibfield  {author} {\bibinfo {author} {\bibfnamefont {D.~I.}\ \bibnamefont
  {Ford}},\ }\href {\doibase 10.1119/1.1986094} {\bibfield  {journal} {\bibinfo
   {journal} {American Journal of Physics}\ }\textbf {\bibinfo {volume} {39}},\
  \bibinfo {pages} {215} (\bibinfo {year} {1971})}\BibitemShut {NoStop}%
\bibitem [{\citenamefont {Aitken}(1939)}]{Aitken39}%
  \BibitemOpen
  \bibfield  {author} {\bibinfo {author} {\bibfnamefont {A.~C.}\ \bibnamefont
  {Aitken}},\ }\href@noop {} {\emph {\bibinfo {title} {Determinants and
  Matrices}}}\ (\bibinfo  {publisher} {Oliver and Boyd},\ \bibinfo {address}
  {Edinburgh and London},\ \bibinfo {year} {1939})\BibitemShut {NoStop}%
\bibitem [{\citenamefont {Stanley}(1999)}]{Stanley99}%
  \BibitemOpen
  \bibfield  {author} {\bibinfo {author} {\bibfnamefont {R.~P.}\ \bibnamefont
  {Stanley}},\ }\href@noop {} {\emph {\bibinfo {title} {Enumerative
  Combinatorics}}},\ Vol.~\bibinfo {volume} {2}\ (\bibinfo  {publisher}
  {Cambridge University Press},\ \bibinfo {address} {Cambridge},\ \bibinfo
  {year} {1999})\BibitemShut {NoStop}%
\bibitem [{\citenamefont {Bargmann}(1961)}]{Bargmann61}%
  \BibitemOpen
  \bibfield  {author} {\bibinfo {author} {\bibfnamefont {V.}~\bibnamefont
  {Bargmann}},\ }\href {\doibase 10.1002/cpa.3160140303} {\bibfield  {journal}
  {\bibinfo  {journal} {Communications on Pure and Applied Mathematics}\
  }\textbf {\bibinfo {volume} {14}},\ \bibinfo {pages} {187} (\bibinfo {year}
  {1961})}\BibitemShut {NoStop}%
\bibitem [{\citenamefont {Bargmann}(1967)}]{Bargmann67}%
  \BibitemOpen
  \bibfield  {author} {\bibinfo {author} {\bibfnamefont {V.}~\bibnamefont
  {Bargmann}},\ }\href {\doibase 10.1002/cpa.3160200102} {\bibfield  {journal}
  {\bibinfo  {journal} {Communications on Pure and Applied Mathematics}\
  }\textbf {\bibinfo {volume} {20}},\ \bibinfo {pages} {1} (\bibinfo {year}
  {1967})}\BibitemShut {NoStop}%
\bibitem [{\citenamefont {Loh}\ \emph {et~al.}(1990)\citenamefont {Loh},
  \citenamefont {Gubernatis}, \citenamefont {Scalettar}, \citenamefont {White},
  \citenamefont {Scalapino},\ and\ \citenamefont {Sugar}}]{Loh90}%
  \BibitemOpen
  \bibfield  {author} {\bibinfo {author} {\bibfnamefont {E.~Y.}\ \bibnamefont
  {Loh}}, \bibinfo {author} {\bibfnamefont {J.~E.}\ \bibnamefont {Gubernatis}},
  \bibinfo {author} {\bibfnamefont {R.~T.}\ \bibnamefont {Scalettar}}, \bibinfo
  {author} {\bibfnamefont {S.~R.}\ \bibnamefont {White}}, \bibinfo {author}
  {\bibfnamefont {D.~J.}\ \bibnamefont {Scalapino}}, \ and\ \bibinfo {author}
  {\bibfnamefont {R.~L.}\ \bibnamefont {Sugar}},\ }\href {\doibase
  10.1103/PhysRevB.41.9301} {\bibfield  {journal} {\bibinfo  {journal} {Phys.
  Rev. B}\ }\textbf {\bibinfo {volume} {41}},\ \bibinfo {pages} {9301}
  (\bibinfo {year} {1990})}\BibitemShut {NoStop}%
\bibitem [{\citenamefont {Reine}\ \emph {et~al.}(2012)\citenamefont {Reine},
  \citenamefont {Helgaker},\ and\ \citenamefont {Lindh}}]{Reine12}%
  \BibitemOpen
  \bibfield  {author} {\bibinfo {author} {\bibfnamefont {S.}~\bibnamefont
  {Reine}}, \bibinfo {author} {\bibfnamefont {T.}~\bibnamefont {Helgaker}}, \
  and\ \bibinfo {author} {\bibfnamefont {R.}~\bibnamefont {Lindh}},\ }\href
  {\doibase 10.1002/wcms.78} {\bibfield  {journal} {\bibinfo  {journal} {Wiley
  Interdisciplinary Reviews: Computational Molecular Science}\ }\textbf
  {\bibinfo {volume} {2}},\ \bibinfo {pages} {290} (\bibinfo {year}
  {2012})}\BibitemShut {NoStop}%
\bibitem [{\citenamefont {Hohenberg}\ and\ \citenamefont
  {Kohn}(1964)}]{Hohenberg64}%
  \BibitemOpen
  \bibfield  {author} {\bibinfo {author} {\bibfnamefont {P.}~\bibnamefont
  {Hohenberg}}\ and\ \bibinfo {author} {\bibfnamefont {W.}~\bibnamefont
  {Kohn}},\ }\href {\doibase 10.1103/PhysRev.136.B864} {\bibfield  {journal}
  {\bibinfo  {journal} {Phys. Rev.}\ }\textbf {\bibinfo {volume} {136}},\
  \bibinfo {pages} {B864} (\bibinfo {year} {1964})}\BibitemShut {NoStop}%
\bibitem [{\citenamefont {Ceperley}(1991)}]{Ceperley91}%
  \BibitemOpen
  \bibfield  {author} {\bibinfo {author} {\bibfnamefont {D.}~\bibnamefont
  {Ceperley}},\ }\href {\doibase 10.1007/BF01030009} {\bibfield  {journal}
  {\bibinfo  {journal} {Journal of Statistical Physics}\ }\textbf {\bibinfo
  {volume} {63}},\ \bibinfo {pages} {1237} (\bibinfo {year}
  {1991})}\BibitemShut {NoStop}%
\bibitem [{\citenamefont {Loos}\ and\ \citenamefont
  {Bressanini}(2015)}]{Loos15}%
  \BibitemOpen
  \bibfield  {author} {\bibinfo {author} {\bibfnamefont {P.-F.}\ \bibnamefont
  {Loos}}\ and\ \bibinfo {author} {\bibfnamefont {D.}~\bibnamefont
  {Bressanini}},\ }\href {\doibase 10.1063/1.4922159} {\bibfield  {journal}
  {\bibinfo  {journal} {The Journal of Chemical Physics}\ }\textbf {\bibinfo
  {volume} {142}},\ \bibinfo {pages} {214112} (\bibinfo {year}
  {2015})}\BibitemShut {NoStop}%
\bibitem [{\citenamefont {Bajdich}\ \emph {et~al.}(2005)\citenamefont
  {Bajdich}, \citenamefont {Mitas}, \citenamefont {Drobn{\'y}},\ and\
  \citenamefont {Wagner}}]{Bajdich05}%
  \BibitemOpen
  \bibfield  {author} {\bibinfo {author} {\bibfnamefont {M.}~\bibnamefont
  {Bajdich}}, \bibinfo {author} {\bibfnamefont {L.}~\bibnamefont {Mitas}},
  \bibinfo {author} {\bibfnamefont {G.}~\bibnamefont {Drobn{\'y}}}, \ and\
  \bibinfo {author} {\bibfnamefont {L.~K.}\ \bibnamefont {Wagner}},\ }\href
  {\doibase 10.1103/PhysRevB.72.075131} {\bibfield  {journal} {\bibinfo
  {journal} {Phys. Rev. B}\ }\textbf {\bibinfo {volume} {72}},\ \bibinfo
  {pages} {075131} (\bibinfo {year} {2005})}\BibitemShut {NoStop}%
\bibitem [{\citenamefont {Stanley}(1978)}]{Stanley78}%
  \BibitemOpen
  \bibfield  {author} {\bibinfo {author} {\bibfnamefont {R.~P.}\ \bibnamefont
  {Stanley}},\ }\href {\doibase 10.1016/0001-8708(78)90045-2} {\bibfield
  {journal} {\bibinfo  {journal} {Advances in Mathematics}\ }\textbf {\bibinfo
  {volume} {28}},\ \bibinfo {pages} {57} (\bibinfo {year} {1978})}\BibitemShut
  {NoStop}%
\bibitem [{\citenamefont {Yamanaka}\ \emph {et~al.}(2005)\citenamefont
  {Yamanaka}, \citenamefont {Koizumi}, \citenamefont {Kitagawa}, \citenamefont
  {Kawakami}, \citenamefont {Okumura},\ and\ \citenamefont
  {Yamaguchi}}]{Yamanaka05}%
  \BibitemOpen
  \bibfield  {author} {\bibinfo {author} {\bibfnamefont {S.}~\bibnamefont
  {Yamanaka}}, \bibinfo {author} {\bibfnamefont {K.}~\bibnamefont {Koizumi}},
  \bibinfo {author} {\bibfnamefont {Y.}~\bibnamefont {Kitagawa}}, \bibinfo
  {author} {\bibfnamefont {T.}~\bibnamefont {Kawakami}}, \bibinfo {author}
  {\bibfnamefont {M.}~\bibnamefont {Okumura}}, \ and\ \bibinfo {author}
  {\bibfnamefont {K.}~\bibnamefont {Yamaguchi}},\ }\href {\doibase
  10.1002/qua.20784} {\bibfield  {journal} {\bibinfo  {journal} {International
  Journal of Quantum Chemistry}\ }\textbf {\bibinfo {volume} {105}},\ \bibinfo
  {pages} {687} (\bibinfo {year} {2005})}\BibitemShut {NoStop}%
\bibitem [{\citenamefont {Bressanini}(2012)}]{Bressanini12}%
  \BibitemOpen
  \bibfield  {author} {\bibinfo {author} {\bibfnamefont {D.}~\bibnamefont
  {Bressanini}},\ }\href {\doibase 10.1103/PhysRevB.86.115120} {\bibfield
  {journal} {\bibinfo  {journal} {Phys. Rev. B}\ }\textbf {\bibinfo {volume}
  {86}},\ \bibinfo {pages} {115120} (\bibinfo {year} {2012})}\BibitemShut
  {NoStop}%
\bibitem [{\citenamefont {Milne}(2015)}]{MilneAG}%
  \BibitemOpen
  \bibfield  {author} {\bibinfo {author} {\bibfnamefont {J.~S.}\ \bibnamefont
  {Milne}},\ }\href@noop {} {\enquote {\bibinfo {title} {Algebraic geometry
  (v6.01)},}\ } (\bibinfo {year} {2015}),\ \bibinfo {note} {available at
  www.jmilne.org/math/}\BibitemShut {NoStop}%
\bibitem [{\citenamefont {Dieudonne}(1972)}]{Dieudonne72}%
  \BibitemOpen
  \bibfield  {author} {\bibinfo {author} {\bibfnamefont {J.}~\bibnamefont
  {Dieudonne}},\ }\href {\doibase 10.2307/2317664} {\bibfield  {journal}
  {\bibinfo  {journal} {The American Mathematical Monthly}\ }\textbf {\bibinfo
  {volume} {79}},\ \bibinfo {pages} {827} (\bibinfo {year} {1972})}\BibitemShut
  {NoStop}%
\bibitem [{\citenamefont {Filinov}(2014)}]{Filinov14}%
  \BibitemOpen
  \bibfield  {author} {\bibinfo {author} {\bibfnamefont {V.}~\bibnamefont
  {Filinov}},\ }\href {\doibase 10.1134/S0018151X14040105} {\bibfield
  {journal} {\bibinfo  {journal} {High Temperature}\ }\textbf {\bibinfo
  {volume} {52}},\ \bibinfo {pages} {615} (\bibinfo {year} {2014})}\BibitemShut
  {NoStop}%
\bibitem [{\citenamefont {Georges}\ \emph {et~al.}(1996)\citenamefont
  {Georges}, \citenamefont {Kotliar}, \citenamefont {Krauth},\ and\
  \citenamefont {Rozenberg}}]{Georges96}%
  \BibitemOpen
  \bibfield  {author} {\bibinfo {author} {\bibfnamefont {A.}~\bibnamefont
  {Georges}}, \bibinfo {author} {\bibfnamefont {G.}~\bibnamefont {Kotliar}},
  \bibinfo {author} {\bibfnamefont {W.}~\bibnamefont {Krauth}}, \ and\ \bibinfo
  {author} {\bibfnamefont {M.~J.}\ \bibnamefont {Rozenberg}},\ }\href {\doibase
  10.1103/RevModPhys.68.13} {\bibfield  {journal} {\bibinfo  {journal} {Rev.
  Mod. Phys.}\ }\textbf {\bibinfo {volume} {68}},\ \bibinfo {pages} {13}
  (\bibinfo {year} {1996})}\BibitemShut {NoStop}%
\bibitem [{\citenamefont {{\v Zivkovi\'c}}\ and\ \citenamefont
  {{Maksi\'c}}(1968)}]{Zivkovic68}%
  \BibitemOpen
  \bibfield  {author} {\bibinfo {author} {\bibfnamefont {T.}~\bibnamefont {{\v
  Zivkovi\'c}}}\ and\ \bibinfo {author} {\bibfnamefont {Z.~B.}\ \bibnamefont
  {{Maksi\'c}}},\ }\href {\doibase 10.1063/1.1670551} {\bibfield  {journal}
  {\bibinfo  {journal} {The Journal of Chemical Physics}\ }\textbf {\bibinfo
  {volume} {49}},\ \bibinfo {pages} {3083} (\bibinfo {year}
  {1968})}\BibitemShut {NoStop}%
\bibitem [{\citenamefont {McMurchie}\ and\ \citenamefont
  {Davidson}(1978)}]{McMurchie78}%
  \BibitemOpen
  \bibfield  {author} {\bibinfo {author} {\bibfnamefont {L.~E.}\ \bibnamefont
  {McMurchie}}\ and\ \bibinfo {author} {\bibfnamefont {E.~R.}\ \bibnamefont
  {Davidson}},\ }\href {\doibase 10.1016/0021-9991(78)90092-X} {\bibfield
  {journal} {\bibinfo  {journal} {Journal of Computational Physics}\ }\textbf
  {\bibinfo {volume} {26}},\ \bibinfo {pages} {218 } (\bibinfo {year}
  {1978})}\BibitemShut {NoStop}%
\end{thebibliography}
\end{document}